\documentclass[a4paper,fleqn,usenatbib]{mnras}
\usepackage{enumerate}
\usepackage[T1]{fontenc}
\usepackage{ae,aecompl}
\usepackage{indentfirst}
\usepackage{graphicx}
\usepackage{amsmath}
\usepackage{amssymb}
\usepackage{color}
\usepackage{mathptmx}
\usepackage{epsfig}
\usepackage{wasysym}
\usepackage{xfrac}
\usepackage{pifont}
\usepackage[table,xcdraw]{xcolor}
\usepackage{booktabs}
\usepackage{multirow}
\usepackage{comment}
\usepackage{subfigure}
\usepackage{longtable}
\usepackage[normalem]{ulem}

\title[Low mass ellipticals form via secular accretion]{Relaxed blue ellipticals: accretion-driven stellar growth is a key evolutionary channel for low mass elliptical galaxies}

\author[I. Lazar et al.]{
I. Lazar,$^{1}$\thanks{E-mail: i.lazar@herts.ac.uk}, S. Kaviraj,$^{1}$ G. Martin,$^{2,3}$ C. Laigle,$^{4}$ A. Watkins$^{1}$ and R. A. Jackson$^{5}$\\
$^{1}$Centre for Astrophysics Research, School of Physics, Astronomy and Mathematics, University of Hertfordshire, College Lane, Hatfield AL10 9AB, UK\\
$^{2}$Korea Astronomy and Space Science Institute, 776 Daedeokdae-ro, Yuseong-gu, Daejeon 34055, Korea\\
$^{3}$Steward Observatory, University of Arizona, 933 N. Cherry Ave, Tucson, AZ, USA\\
$^{4}$Institut d’Astrophysique de Paris, UMR 7095, CNRS, and Sorbonne Université, 98 bis boulevard Arago, 75014 Paris, France\\
$^{5}$Department of Physics and Astronomy, University of Victoria, Victoria, BC, Canada V8P 5C2}

\pubyear{2023}

\begin{document}
\label{firstpage}
\pagerange{\pageref{firstpage}--\pageref{lastpage}}
\maketitle

\begin{abstract}
How elliptical galaxies form is a key question in observational cosmology. While the formation of massive ellipticals is strongly linked to mergers, the low mass (M$_{\star}$/M$_{\odot}$ < 10$^{9.5}$) regime remains less well explored. In particular, studying elliptical populations when they are blue, and therefore rapidly building stellar mass, offers strong constraints on their formation. Here, we study 108 blue, low-mass ellipticals (which have a median stellar mass of 10$^{8.7}$ M$_{\odot}$) at $z<0.3$ in the COSMOS field. Visual inspection of extremely deep optical HSC images indicates that less than 3 per cent of these systems have visible tidal features, a factor of 2 less than the incidence of tidal features in a control sample of galaxies with the same distribution of stellar mass and redshift. This suggests that the star formation activity in these objects is not driven by mergers or interactions but by secular gas accretion. We combine accurate physical parameters from the COSMOS2020 catalog, with measurements of local density and the locations of galaxies in the cosmic web, to show that our blue ellipticals reside in low-density environments, further away from nodes and large-scale filaments than other galaxies. At similar stellar masses and environments, blue ellipticals outnumber their normal (red) counterparts by a factor of 2. Thus, these systems are likely progenitors of not only normal ellipticals at similar stellar mass but, given their high star formation rates, also of ellipticals at higher stellar masses. Secular gas accretion, therefore, likely plays a significant (and possibly dominant) role in the stellar assembly of elliptical galaxies in the low mass regime.

\end{abstract}




\section{Introduction}
Our currently-accepted paradigm of structure formation postulates a hierarchical build up of stellar mass, driven by interactions and merging \citep[e.g ][]{Press1974,White1978,Blumenthal1984}. Within this picture, elliptical galaxies, at least those with high stellar masses, are considered to be the natural dynamical end-points of galaxy evolution, intimately linked to galaxy merging \citep[e.g.][]{Dressler1980,Moore1996,Dekel2006,Faber2007}. The strong gravitational torques created during galaxy interactions can destroy discs, by removing angular momentum and randomising stellar orbits, thus producing the smooth, relatively featureless morphologies typically found in elliptical systems \citep[e.g.][]{Martin2018a, Park2019}.

Various pieces of evidence appear to support this picture in the high-mass regime. Both theoretical and observational studies suggest an increase in the fraction of massive elliptical systems in regions of higher density, where the frequency of merging is expected to be higher (e.g. the central regions of clusters), with the morphological mix being increasingly dominated by late-type (i.e. disc-dominated) galaxies as local density decreases \citep[e.g.][]{Dressler1980, Brough2006, Hopkins2010, Li2012, Deeley2017, Sengupta2022, Kolokythas2022}. The tendency for merging to create, or at least reinforce, the spheroidal elements of a massive galaxy appears consistent with a large variety of morphological and kinematical phenomena that are present in many nearby ($z<1$) massive ellipticals \citep{Quilley2022} and which are signposts of galaxy interactions, e.g. tidal features \citep{Toomre1972, Peirani2010, Kaviraj2014b, Kaviraj2019}, shells \citep{Quinn1984, Cooper2011, Bilek2022}, stellar streams \citep{Johnston1999} or kinematically decoupled cores \citep{Jedrzejewski1988, Miller1990, Forbes1995, Cappellari2011, Bois2011, Guerou2015}.

While the dynamical changes in the structure of today's massive ellipticals appears to have taken place via merging over cosmic time \citep[e.g.][]{Bender1988,Boylan-Kolchin2006,Martin2018a}, the bulk of the stellar mass in these galaxies appears to have formed relatively rapidly at high redshift, as evidenced by their red optical colours \citep[e.g][]{Larson1974, Chiosi2002, Ferreras2005, Baldry2007, Kaviraj2007, Bamford2009}, which exhibit little scatter \citep[e.g.][]{Stanford1998, Dokkum2000} and high alpha-to-iron ratios \citep[e.g.][]{Thomas2005}. 

Both theory \citep[e.g.][]{Martin2019} and observations \citep[e.g.][]{Lofthouse2017} suggest, however, that the majority ($\sim$80 per cent) of the old stellar populations that form in the early Universe and dominate massive ellipticals at the present day form through gas accretion, rather than mergers. This appears to be driven both by the relative paucity of merger events over cosmic time \citep[particularly major mergers, e.g.][]{Conselice2003} and by the fact that the early Universe is extremely gas rich and mergers do not enhance star formation rates beyond what is possible by gas accretion alone \citep[e.g.][]{Kaviraj2013,Lofthouse2017,Fensch2017}. Stellar assembly continues in massive ellipticals over cosmic time \citep[e.g.][]{Fukugita2004, Kannappan2009, Yi2005, Kaviraj2007, Kelvin2014}, with a significant minority (20 -- 30 per cent) of stellar mass forming at $z<1$ \citep{Kaviraj2008}. At late epochs, the star formation activity in massive ellipticals appears to be more influenced by merging \citep[e.g.][]{George2017, Omar2018}, with most of the interactions being unequal mass ratio (i.e. minor) mergers \citep[e.g.][]{Kaviraj2011}.

While massive ellipticals have been extensively studied, both via observations and theory, much less is known about the evolutionary pathways of their low mass counterparts. This is especially true for these populations outside the very nearby Universe, because past large surveys (like the SDSS) offer large footprints but are relatively shallow, making it more difficult to detect lower luminosity objects outside the very local Universe \citep[e.g.][]{Jackson2021,Davis2022}. Nevertheless, a growing literature has recently probed the evolution of local ellipticals in the low mass (M$_{\star}$ < 10$^{9.5}$ M$_\odot$) regime. Many of these studies have focussed on blue ellipticals which are brighter and easier to detect in surveys like the SDSS. Apart from their increased detectability, blue ellipticals are also a particularly efficient route to understanding the processes that dominate the stellar assembly of these systems, since they are in a phase where they are actively building significant fractions of their stellar mass.  

These studies have postulated several mechanisms that may contribute to the formation of blue low-mass ellipticals, including secular gas accretion \citep[e.g.][]{Mahajan2018}, major and minor mergers \citep[e.g.][]{Schawinski2009a,Chung2019} or a combination of both \citep[e.g.][]{Moffett2019}. It is worth noting that the unavailability, in past studies, of deep images, that are required to reveal merger-driven tidal features \citep[e.g.][]{Kaviraj2014b}, has been a significant impediment to conclusively probing the role of mergers in the formation of these objects. Since blue ellipticals will become members of the general (red) elliptical population when they quench \citep[e.g.][]{Meyer2014}, these galaxies are valuable as progenitor systems which can provide insights into the processes that form the low mass elliptical population as a whole.

Here, we study a large sample of low-redshift ($z<0.3$) blue elliptical galaxies, which have low stellar masses, with a median stellar mass of 10$^{8.7}$ M$_\odot$. We explore the physical processes that drive their stellar assembly, using multi-wavelength data in the COSMOS field, which includes extremely deep imaging from the Hyper Suprime-Cam Subaru Strategic Program \citep[][; HSC-SSP]{Aihara2022}. Our study offers several novelties which allow us to select blue ellipticals, probe their properties and constrain their link to normal ellipticals with higher precision and in greater detail than has been possible in previous work. 

Our blue ellipticals are selected from HSC-SSP images using a novel unsupervised machine-learning algorithm, which is designed to autonomously separate objects with similar morphological properties (e.g. colour, texture and luminosity) in large survey datasets. \textcolor{black}{As described below, this automated selection is then augmented by Sersic fitting to remove stars and ensure that the objects being selected have the strongly peaked profiles expected in ellipticals. The HSC-SSP images used in this study have point-source detection limits of 28.1, 27.7 and 27.4 magnitudes in the $g$, $r$ and $i$ bands respectively, around 6 magnitudes deeper than standard depth imaging from the SDSS}. These represent the deepest survey data used for the analysis of such blue ellipticals to date. This allows us to probe the role of mergers in driving the formation of our blue ellipticals, by looking for merger-induced tidal features more effectively than has been possible in previous work. 

The exquisite deep, multi-wavelength (UV to infrared) photometric data available in the central regions of the HSC-SSP COSMOS field results in extremely precise photometric redshifts (with accuracies better than 1 per cent for massive galaxies) and physical parameters such as stellar masses and star formation rates (SFRs), available via the COSMOS2020 catalog \citep{Weaver2022}. Finally, the accurate redshifts in this field enable us to use the DisPerSE \citep{Sousbie2011} filament-finding algorithm to construct density maps and the filamentary structure of the cosmic web. This is then used to study the local density of our blue ellipticals and their topological locations (i.e. proximity to nodes, large-scale filaments and voids) with higher precision than has been possible in past studies.   

This paper is structured as follows. In Section \ref{sec:algorithm} we describe the machine-learning algorithm used to extract our blue ellipticals and the properties of the HSC-SSP on which the algorithm has been implemented. In Section \ref{sec:cosmos2020} we describe the COSMOS2020 catalog from which we extract photometric redshifts and physical parameters for our galaxies. In Section \ref{sec:disperse} we describe the algorithm used to construct maps of local density and topological structure that we use to study the locations of our galaxies in the cosmic web. In Section \ref{sec:analysis} we explore the role of different processes (e.g. mergers and gas accretion) in the stellar assembly of our blue ellipticals and explore their connection to normal ellipticals (i.e. those that are not blue) to make a broad comment on how the bulk of the stellar mass in low mass elliptical galaxies is likely to form. We summarise our findings in Section \ref{sec:summary}. 


\section{Galaxy morphologies via unsupervised machine learning}
\label{sec:algorithm}

Our blue ellipticals are identified using an unsupervised machine learning (UML) algorithm (Lazar et al. in prep), which has previously been implemented on images from both the \textit{Hubble Space Telescope} \citep{Hocking2018} and the HSC-SSP \citep{Martin2020}. The algorithm samples a large number of random patches (8 by 8 pixels in size) in the footprints of the objects of interest (in our case, galaxies), in multi-band survey imaging. Here, we use HSC-SSP images of galaxies in the \textit{griz} bands. The algorithm then uses a radial Fourier transform to construct a `feature vector' for each patch, that holds information about its physical properties (principally colour, texture and intensity). Patches are clustered via a growing neural gas algorithm \citep[][]{Fritzke1995} and hierarchical clustering \citep{Johnson1967}, and galaxies with similar patch properties are then grouped together using k-means clustering. Arbitrarily large galaxy samples are rapidly and autonomously compressed into $\sim$150 `morphological clusters', with typical purities greater than 90 per cent. $\sim$150 clusters are needed because identical morphologies at different redshifts can occupy separate clusters (since galaxies change in size with redshift) and because this allows for a better morphological diversity between different clusters.

\begin{figure}
    \centering
    \includegraphics[width=0.45\textwidth]{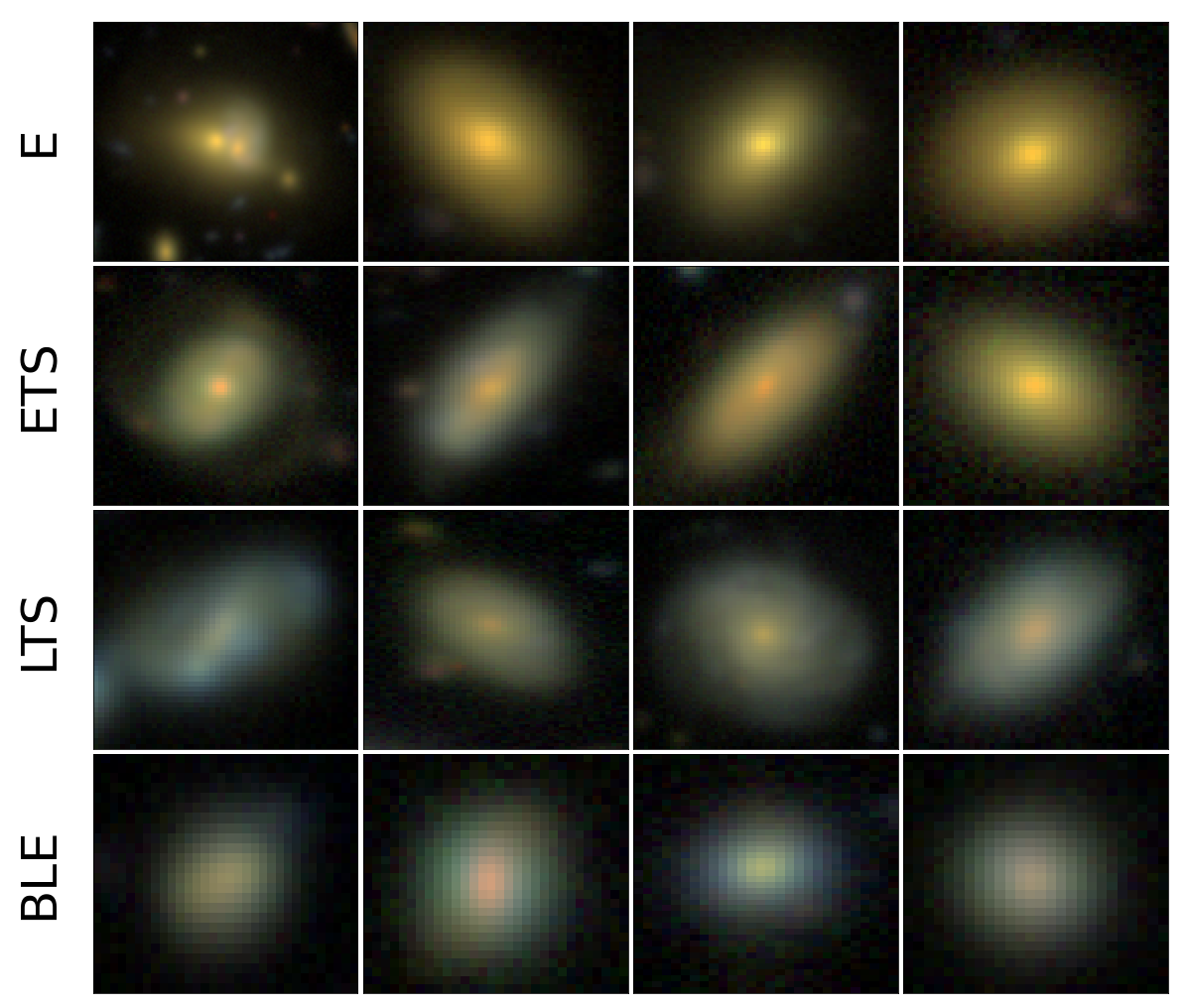}
    \caption{Examples of morphological clusters (show in different rows), outputted by the unsupervised machine learning algorithm, which have been visually associated with Hubble types: ellipticals (E), early type spirals (ETS), late type spirals (LTS) and blue ellipticals (BLE).}
    \label{fig:gals}
\end{figure}

Each of these clusters is then associated, via visual inspection, with standard Hubble types. To perform this labelling, we visually inspect a sample of 30 objects from each cluster, whose feature vectors have varying Euclidean distances from the cluster centre in feature space - we select 10 which are closest to the centre, 10 which are furthest and 10 random objects in between. We label clusters into broad Hubble types: ellipticals, blue ellipticals and early- and late-type spirals (rows 1 - 4 in Figure \ref{fig:gals}). As described in \citet{Martin2020}, these clusters obey known morphological trends in the literature e.g. stellar mass functions and the relationships between stellar mass, SFR and colour as a function of morphology. Peculiar objects such as blue ellipticals (row 4 in Figure \ref{fig:gals}) are naturally segregated into separate morphological clusters. We refer readers to \citet{Martin2020} for more details of the algorithm and note that morphological catalogs for the entire HSC-SSP Public Data Release 3 (PDR3) will be released shortly in a forthcoming paper (Lazar in prep.).

Note that, prior to inputting galaxy footprints into our UML classification algorithm, objects that are likely to be stars are initially filtered out by only selecting objects which have an HSC \texttt{extendedness} flag set to 1 (indicating that they are extended objects and not stars) and which have the \texttt{type} parameter in the COSMOS2020 catalog set to 0 (indicating, again, that they are galaxies). We perform additional filtering for stars when we select our final sample of blue ellipticals, as described in Section \ref{sec:identification}. Note also that we focus, in this particular study, on the nearby Universe ($z<0.3$), where galaxies are extended, redshift errors are small and surface-brightness dimming is minimised, which aids the visual identification of tidal features that are signposts for recent mergers and interactions.

\section{COSMOS2020: accurate physical parameters for galaxies}
\label{sec:cosmos2020}

We obtain physical parameters for our galaxies (photometric redshifts, stellar masses, rest-frame colours and SFRs) from the  COSMOS2020 catalog \citep{Weaver2022}. COSMOS2020 is a high precision value-added catalog of over 1.7 million sources in the central region of the COSMOS field \citep{Scoville2007}. The physical parameters are estimated using photometry in 40 broad and medium band filters, from deep multi-band images taken using the following instruments: GALEX, MegaCam, ACS/HST, Hyper Suprime-Cam, Suprime-Cam, VIRCAM and IRAC. Here, we use parameters derived using the `Classic' method which uses the \texttt{SExtractor} and \texttt{IRACLEAN} codes, with fluxes extracted within circular apertures, after the survey images are homogenized to a common point-spread function. The physical parameters are calculated using the \texttt{LePhare} SED-fitting algorithm \citep{Ilbert2013}. The photometric redshifts have accuracies better than 1 and 4 per cent for bright ($i<22.5$ mag) and faint ($25<i<27$ mag) galaxies respectively, making this catalog ideally suited for our purposes. While accurate photometric redshifts are available from COSMOS2020 for all objects, where possible we use spectroscopic redshifts from the literature (which have been compiled and released as part of the HSC-SSP DR3).\\


\section{Measurement of local density and topological structure}
\label{sec:disperse}

We use the \texttt{DisPerSE} algorithm \citep{Sousbie2011} to measure local density and the locations of galaxies in the cosmic web e.g. their proximity to nodes and large-scale filaments. \texttt{DisPerSE} measures the density field, derived using Delaunay tessellations computed from galaxy positions \citep{Schaap2000}, to connect topological saddle points with peaks (i.e. local maxima or nodes) in the density map via connecting segments, forming a set of ridges which constitute the `skeleton'. The `skeleton' describes the network of large-scale filaments that define the cosmic web, with the stationary points in density maps calculated by \texttt{DisPerSE} (minima, maxima and saddles) denoting the locations of voids, nodes and the centres of large-scale filaments. We refer readers to \citet{Sousbie2011} for further details about the algorithm. 

The properties of the skeleton are determined by a `persistence' parameter, which is used to set a threshold value for defining critical pairs within the density map. A persistence of `\textit{N}' creates a skeleton where all critical pairs with Poisson probabilities below \textit{N}$\sigma$ from the mean are removed. \textcolor{black}{For our density analysis, we follow the methodology of \citet{Laigle2018}, who have implemented \texttt{DisPerSE} on redshift slices of similar widths, constructed from the COSMOS2015 \citep{Laigle2016} catalog, where the thicknesses of the slices are driven by the redshift uncertainties of the galaxy sample. They have also used the Horizon-AGN cosmological simulation \citep{Dubois2014,Kaviraj2017} to show that datasets like COSMOS2015, which offer high photometric redshift precision (and those like COSMOS2020 which provide  higher precision), can retrieve the broad 3D properties of the cosmic web from 2D projected density maps.} We use a persistence of 2, which removes ridges close to the noise level, where structures are likely to be spurious. 

\begin{figure}
    \centering
    \includegraphics[width=0.45\textwidth]{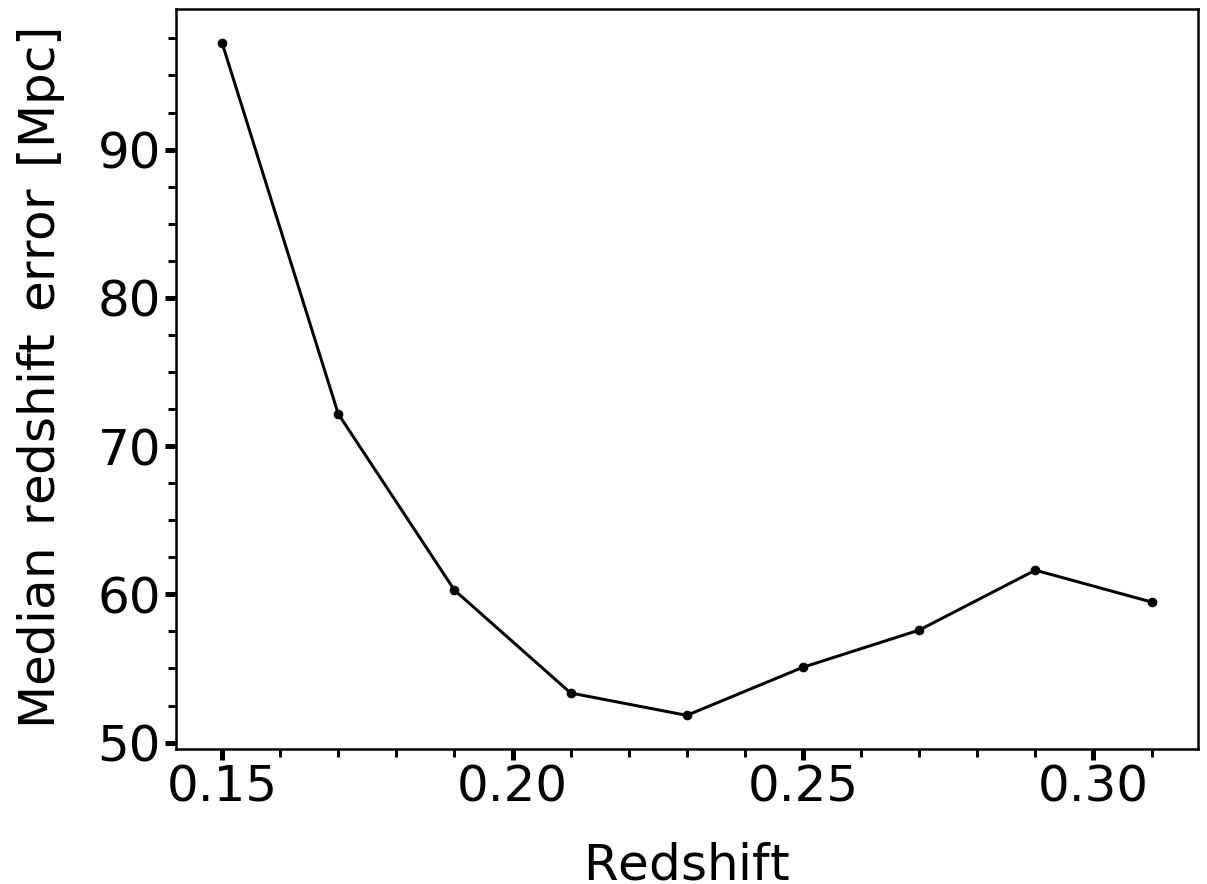}
    \caption{Median redshift error of massive (M$_{\star}$ > 10$^{10}$ M$_\odot$) galaxies converted to a physical distance in Mpc, using the standard cosmology, as a function of redshift. As noted in the text, only galaxies with stellar masses greater than 10$^{10}$ M$_\odot$ are used to construct density maps, as they dominate the local gravitational potential well and have the smallest redshift errors.}
    \label{fig:slice_width}
\end{figure}

The highly accurate redshifts in the COSMOS2020 catalog enable us to use well-defined and relatively narrow redshift slices to build density maps. We only use massive (M$_{\star}$ > 10$^{10}$ M$_{\odot}$) galaxies to build these maps, as they dominate the local gravitational potential well and have the smallest redshift errors. Figure \ref{fig:slice_width} indicates the physical distance associated with the median redshift error of massive galaxies, as a function of redshift. The redshift error reaches a broad minimum in the redshift range $0.18<z<0.3$, corresponding to a physical distance of $\sim$55 -- 65 Mpc. In Section \ref{sec:analysis} we restrict the analysis of the environmental properties of our galaxies to this redshift range.

\begin{figure}
    \centering
    \includegraphics[width=0.45\textwidth]{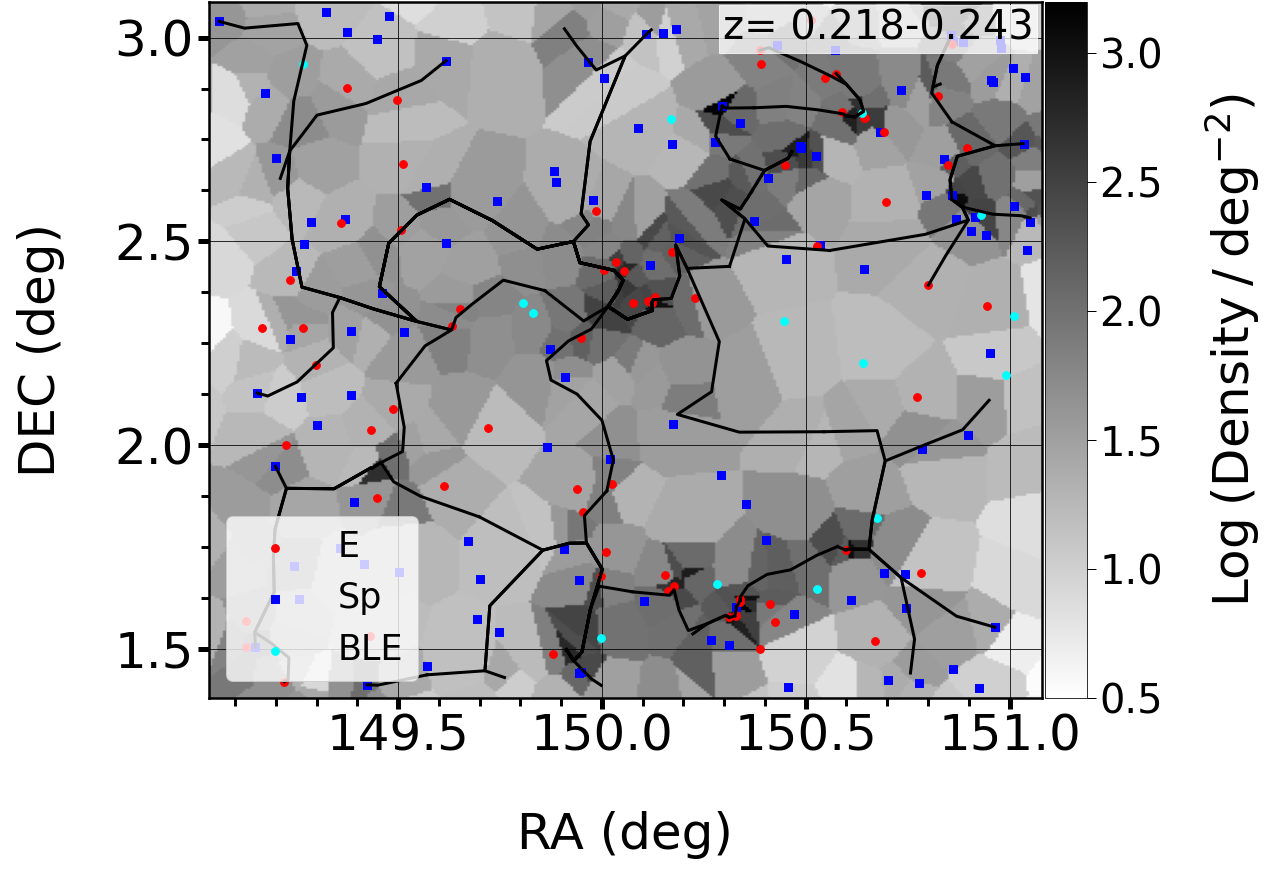}
    \caption{Example density map, with the associated skeleton calculated using \texttt{DisPerSE} shown using the black solid lines. Different morphological classes (red: ellipticals, blue: spirals, cyan: blue ellipticals) are shown overlaid. This particular density map corresponds to the redshift range $0.218 < z < 0.243$. All other density maps used in this study are shown in Appendix \ref{app:dens_maps}.}
    \label{fig:EXMPslice}
\end{figure}

Following \citet{Laigle2018}, we define the slice width to be twice the median redshift error of the massive galaxies, which results in slice widths of $\sim$ 100 Mpc in our redshift range of interest ($0.18<z<0.3$). We produce 9 density slices, each overlapping by 50 Mpc. When constructing each density map, every galaxy is weighted by the area under its redshift probability density function that is contained within the redshift limits of the slice in question. This takes into account the fact that the photometric redshifts of massive galaxies, although very accurate in COSMOS2020, do have associated errors. Figure \ref{fig:EXMPslice} shows an example redshift slice for the reshift range $0.218<z<0.243$, with the associated density map and skeleton calculated using \texttt{DisPerSE}, and galaxies of various morphologies identified using our UML algorithm overplotted. Other redshift slices used in this study are shown in Appendix \ref{app:dens_maps}. 

It is instructive to consider the large-scale structure that is enclosed within the COSMOS2020 region. The COSMOS2020 field has approximate dimensions of 1.6 $\times$ 2 degrees (Figure \ref{fig:EXMPslice}). At the lower ($z\sim0.18$) and upper ($z\sim0.3$) ends of the redshift range we probe here, 1.8 degrees corresponds to a linear size of $\sim$20 and $\sim$30 Mpc respectively. To get insights into the types of structures that might be present in a field of this size, we consider the NewHorizon cosmological simulation \citep{Dubois2021}, which has a similar size to the COSMOS2020 field at these redshifts. The largest dark matter halo in NewHorizon has a mass close to that of the Fornax cluster. It is, likely, therefore that the large scale structure within the COSMOS2020 field contains small clusters like Fornax, along with smaller groups of galaxies.

\section{blue ellipticals: the importance of secular gas accretion and the insignificance of mergers}
\label{sec:analysis}

In this section, we first describe how blue ellipticals are identified using the morphological clusters produced by the UML algorithm described in Section \ref{sec:algorithm}. We then combine their physical and morphological properties with their locations in the cosmic web to explore the processes that are likely to give rise to these systems. Finally, we consider their connection to the general elliptical population and study the implications of our findings on the evolution of low mass elliptical galaxies as a whole.


\subsection{Identification}
\label{sec:identification}

As noted in Section \ref{sec:algorithm}, morphological clusters outputted by our classification algorithm are associated via visual inspection with standard Hubble types - broadly ellipticals and spirals. Elliptical galaxies which are blue naturally end up in their own morphological clusters, as the classification algorithm uses colour as one of its primary discriminants for separating objects of different morphological types. Galaxies in these clusters represent a sample of blue elliptical \textit{candidates} which form the starting point for selecting our final sample of galaxies (amounting to $\sim$1100 initial candidates), as described below. For these candidate systems we (1) perform further filtering for stars and (2) select a subset of objects that have strongly peaked profiles that are typical of elliptical morphologies \citep[e.g.][]{Vaucouleurs1948,Kormendy1977}. This subset forms the final sample of blue ellipticals that underpins our study.

\begin{figure}
    \centering
    \includegraphics[width=0.45\textwidth]{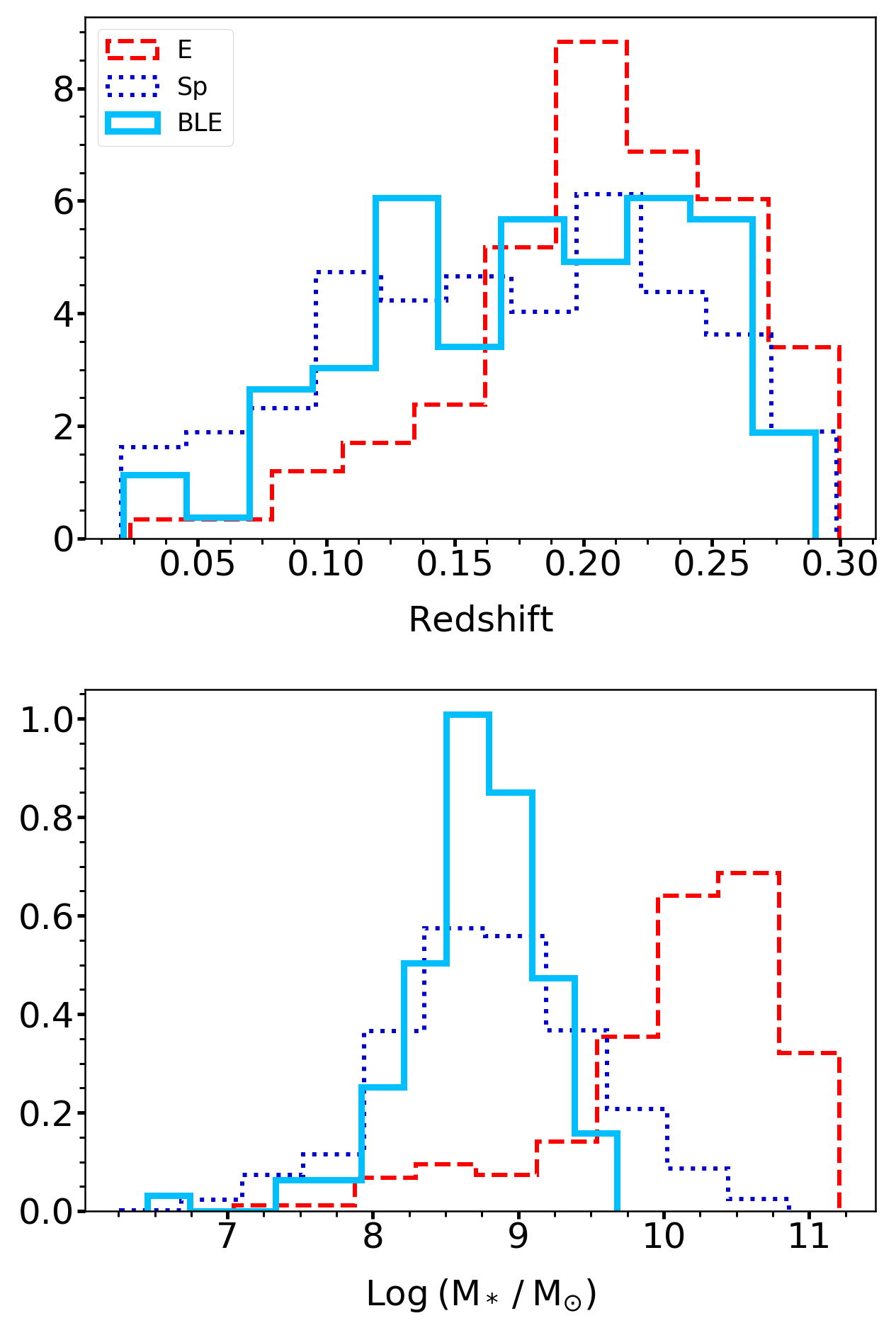}
    \caption{Redshift (top) and stellar mass (bottom) distribution for our blue ellipticals (BLE) and other morphological types: ellipticals (E) and spirals (Sp).}
    \label{fig:mass_redshift}
\end{figure}

\begin{figure}
    \centering
    \includegraphics[width=0.45\textwidth]{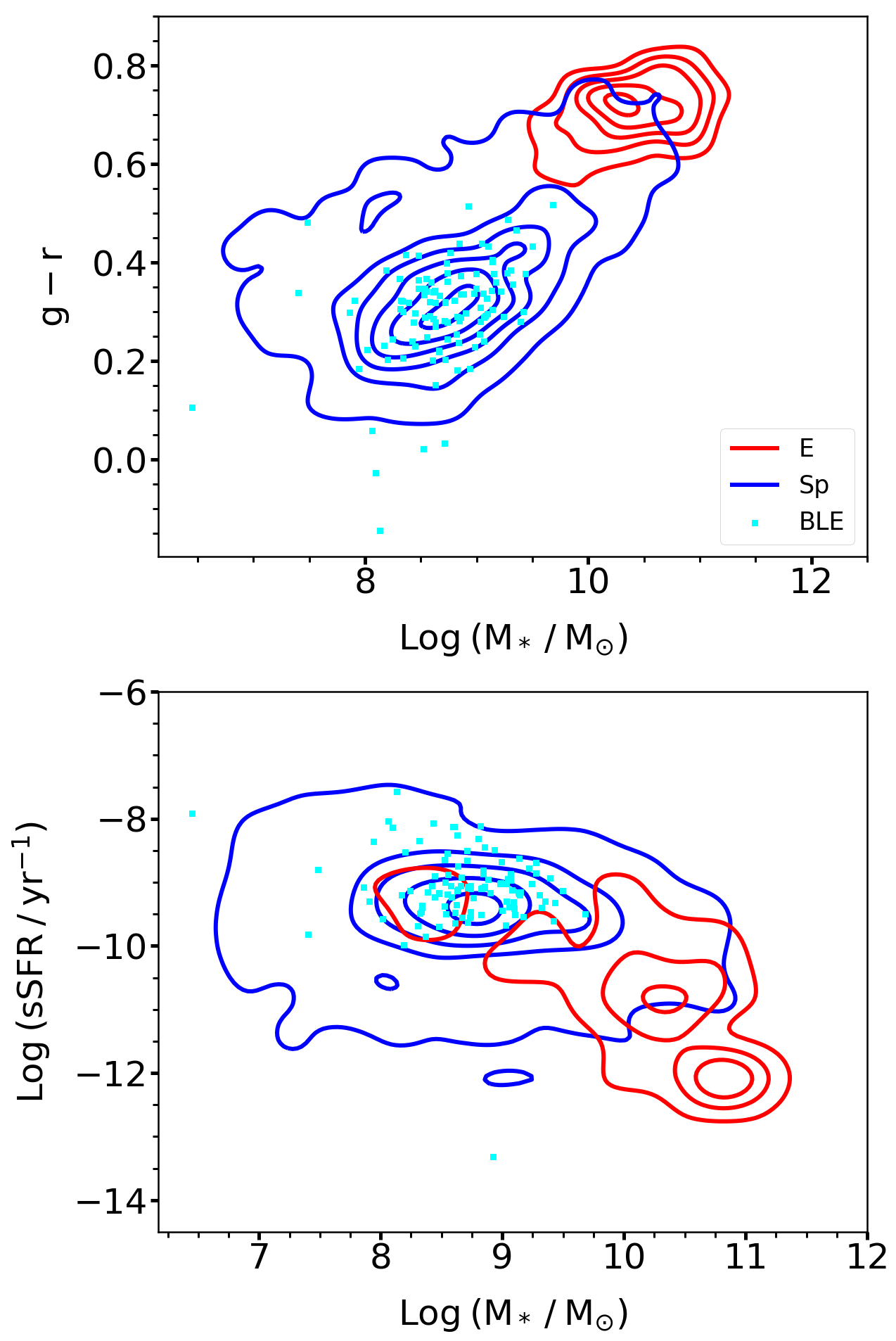}
    \caption{Rest-frame $g-r$ colour (top) and specific star formation rates (bottom) of our blue ellipticals (BLE) and other morphological types: ellipticals (E) and spirals (Sp).}
    \label{fig:colorMass}
\end{figure}

 We first fit the \textit{g} band HSC image of each blue elliptical candidate with a single-component Sersic profile using \texttt{PyImfit} \citep[a Python wrapper for \texttt{Imfit} from][]{Erwin2015}, \textcolor{black}{convolved with the point spread function}. We then remove objects which have Sersic indices less than 4 \citep[since these may contain a disc component, e.g.][]{Vaucouleurs1959,Vaucouleurs1977,Blanton2003,Pannella2006,Propris2016}. Although we perform an initial filtering for stars using two catalog parameters at the outset, before any galaxies are inputted into the classification algorithm (as described above), we take additional steps to ensure that our blue elliptical sample is not contaminated by stars. \textcolor{black}{Note that 40 objects have spectroscopic redshifts, confirming that they are galaxies, and are therefore excluded from the filtering process described below.} 
 
 To perform additional filtering for stars, we use the catalog of \citet{Bailer2019}, using the maximum value from the residual map of the Sersic fits in the $g$-band. The \citet{Bailer2019} catalog contains star, galaxy and quasar classifications for objects in the GAIA Data Release 2 (DR2). The authors make use of a supervised classifier, based on Gaussian Mixture Models, which use the GAIA DR2, cross-matched with spectroscopic classifications from the SDSS as a training set. 
 
 The maximum residual values of objects that are identified as stars and quasars in \citet{Bailer2019} are significantly higher than those for objects that are identified as galaxies. We define a conservative threshold of 1.5 for this maximum residual value, in order to differentiate between stars and galaxies. This threshold value corresponds to the lowest maximum residual value among all objects labelled as stars in the \cite{Bailer2019} catalog. Appendix \ref{app:sersic} shows examples of blue elliptical candidates which have Sersic indices greater than 4 and objects classified as stars by \citet{Bailer2019}. This leaves a final sample of 108 blue ellipticals which forms the basis of our study (of which 40 i.e. $\sim$37 per cent have spectroscopic redshifts).

It is worth noting here that the Sersic index ($n$) threshold which defines elliptical galaxies may vary as a function of stellar mass, with studies in the very nearby Universe \citep[e.g. S$^4$G sample][]{Watkins2022} suggesting that a lower value than $n \sim 4$ may be appropriate in the low mass regime. Nevertheless, we use $n \sim 4$ as our threshold here, as the S$^4$G galaxies probe significantly lower redshifts than our objects and because statistical morphological studies of large samples of low mass galaxies outside the local Universe, which would require deep-wide surveys, are still largely missing. The existence of low mass ellipticals with high concentrations is therefore not ruled out. It is also worth noting that the high central concentrations in our blue ellipticals are, at least in part, driven by central ongoing star formation in these galaxies. The profiles are likely to become shallower as the star formation subsides and the galaxies quench.


\subsection{Physical properties}
\label{sec:prop}

The top and bottom panels of Figure \ref{fig:mass_redshift} show the redshift and stellar mass distributions respectively of our blue ellipticals and other morphological classes in our galaxy sample. Our blue ellipticals have a median stellar mass of 10$^{8.7}$ M$_\odot$ \citep[similar to blue ellipticals studied in some past work e.g.][]{Kannappan2009,Mahajan2018, Moffett2019}. The median redshift of our sample is 0.19. Our blue ellipticals exhibit specific star formation rates and rest-frame colours that are comparable to objects on the main sequence of star-forming galaxies, which is dominated by late-type (i.e. spiral) systems (Figure \ref{fig:colorMass}). As expected, the population of normal elliptical galaxies, in contrast, dominates the red sequence. The behaviour seen here is similar to that in \citet{Schawinski2007} and \citet{Schawinski2014} who have also studied low-mass blue ellipticals in the nearby Universe.

\begin{figure*}
    \centering
    \includegraphics[width=\textwidth]{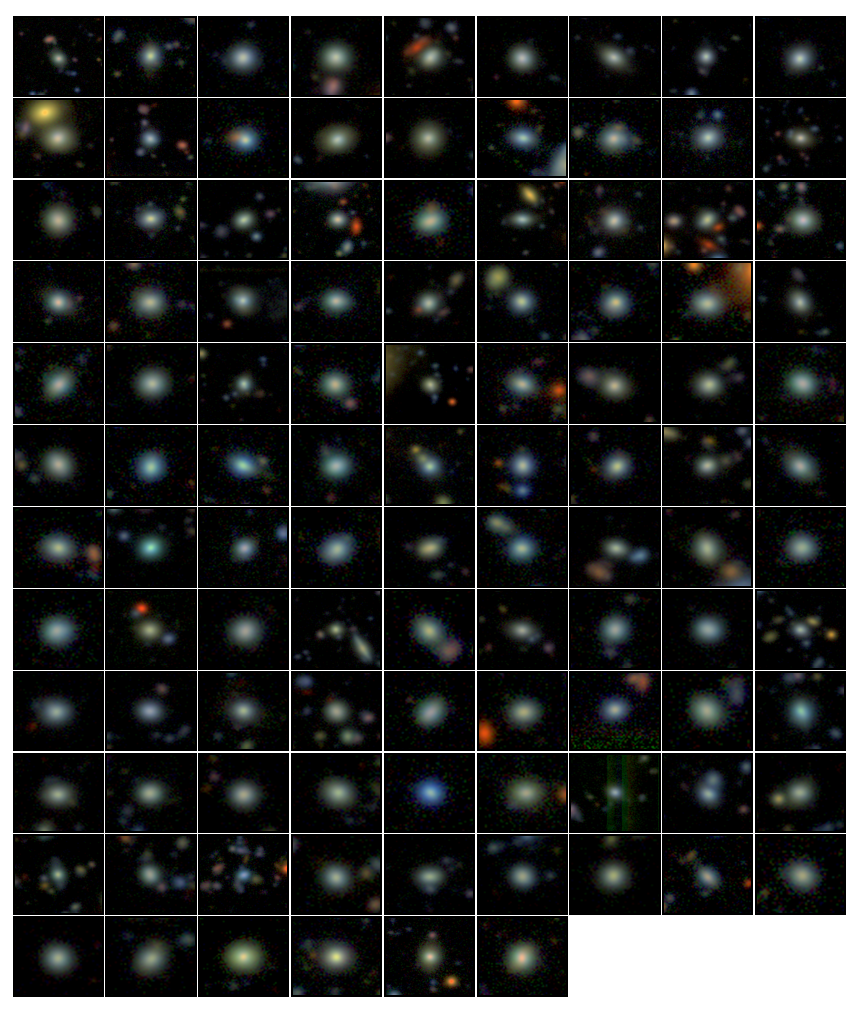}
    \caption{Blue ellipticals which are relaxed i.e. show no evidence for tidal features in the deep HSC images.}
    \label{fig:gals2}
\end{figure*}

\begin{figure}
    \centering
    \includegraphics[width=0.45\textwidth]{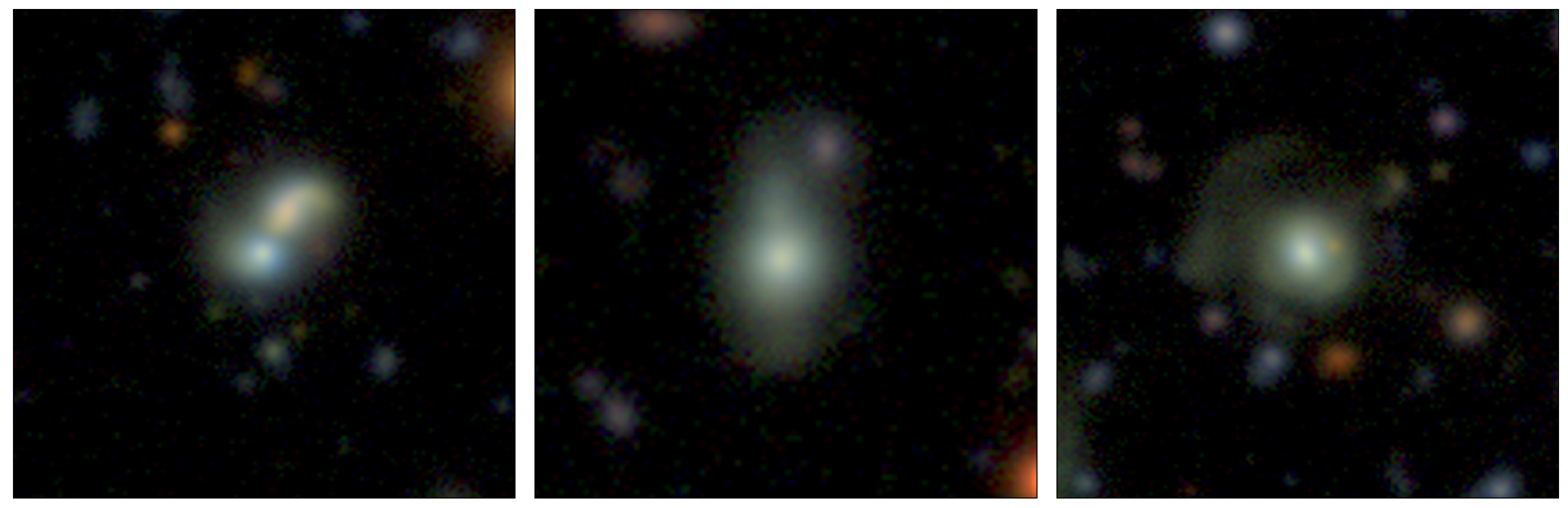}
    \caption{Blue ellipticals which exhibit either evidence of an ongoing merger (left-hand panel) or tidal features indicative of a recent interaction (other panels).}
    \label{fig:gals3}
\end{figure}

\begin{figure}
    \centering
    \includegraphics[width=0.45\textwidth]{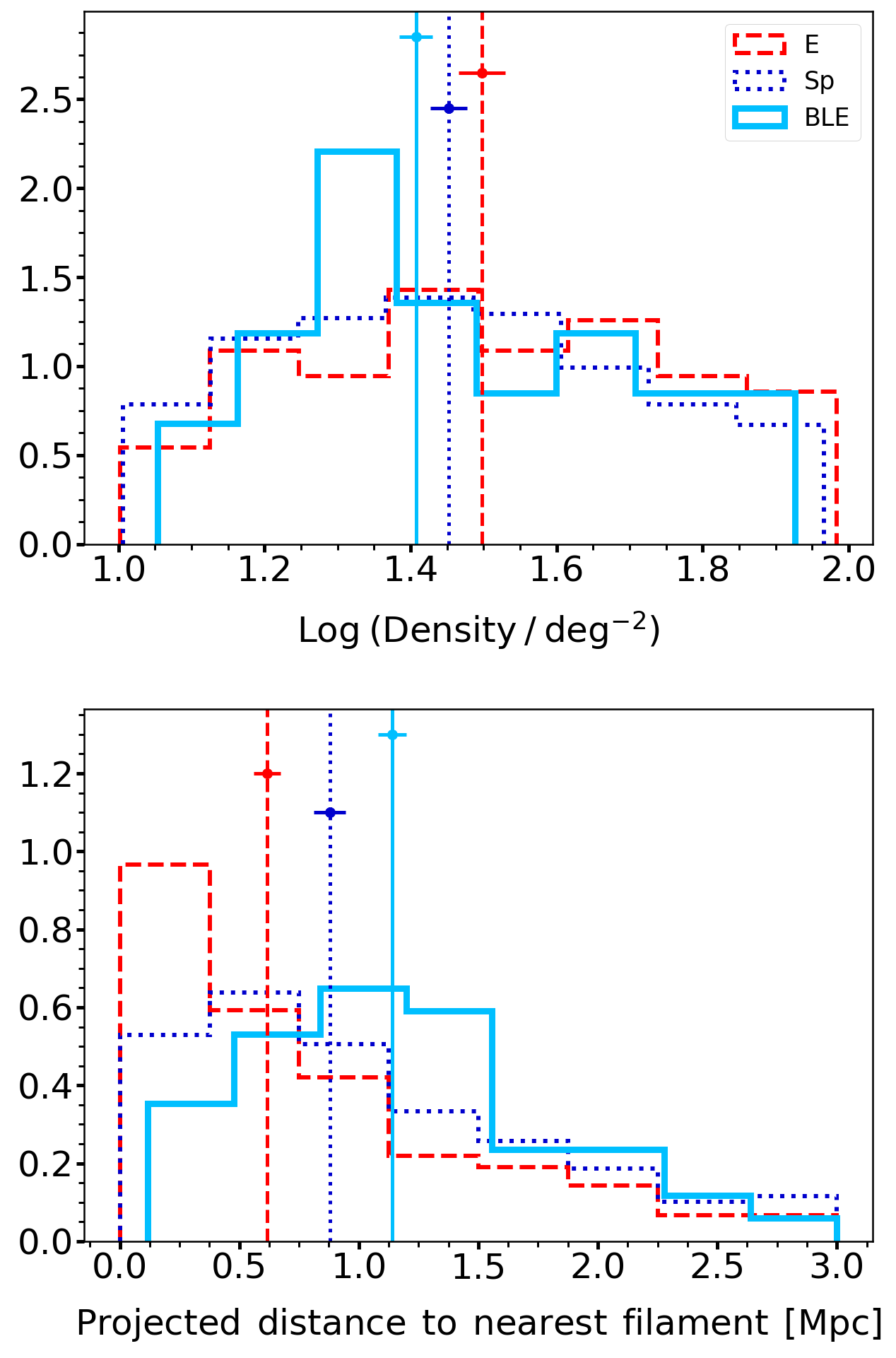}
    \caption{The distributions of local  density (top) and projected distance from the nearest filament (bottom) for different morphological types (red: ellipticals, dark blue : spirals, light blue: blue ellipticals). The medians of the distributions are represented by dashed lines. The standard error in these median values are calculated using 100 bootstrapped iterations and are shown using the solid horizontal lines.}
    \label{fig:densVSfilDist}
\end{figure}


\subsection{Absence of tidal features - blue ellipticals typically do not evolve via galaxy interactions}
\label{sec:img}

We first consider whether the star formation seen in our blue ellipticals is likely to be driven by a recent merger or interaction. The HSC images of our blue ellipticals offer an unprecedented combination of depth and resolution, compared to past surveys like the SDSS (on which many previous studies of blue ellipticals have been based). Recent mergers or interactions produce tidal features \citep[e.g.][]{Toomre1972,Schweizer1982,Mihos1996,Springel2005,Yoon2022,Martin2022}, which are best identified by direct visual inspection of galaxy images \citep[e.g.][]{Darg2010}. Indeed, tidal features are readily detected in images from surveys like the SDSS Stripe 82 \citep[e.g.][]{Kaviraj2014b}, which are around 3 magnitudes shallower than the images we use here. 

Visual inspection of the HSC images indicates that only 3 out of the 108 blue ellipticals (2.8$^{+ 2.6}_{-0.9}$ per cent) show either a tidal feature indicative of a recent interaction or an ongoing merger. To study how this interaction fraction compares to that in the general galaxy population, we select a random sample of 108 galaxies within the COSMOS field which have the same stellar mass and redshift distribution as our blue ellipticals. We then visually inspect the HSC images of this general sample of galaxies. The tidal feature fraction in this general sample is 5.6$^{+3.1}_{-1.5}$ per cent. The errors in the tidal feature fractions have been calculated following \citet{Cameron2011}\footnote{\citet{Cameron2011} calculate accurate Bayesian binomial confidence intervals using the quantiles of the beta distribution. These are considered to be more accurate than simpler methods, like using the normal approximation, which misrepresents the statistical uncertainty under the sampling conditions (e.g. small number counts) that are often encountered in astronomical surveys.}. Our blue ellipticals therefore show a lower tidal feature fraction than the general galaxy population, indicating that mergers are not a preferential channel for their recent evolution.\footnote{It is worth noting that tidal features are also readily visible, in these deep HSC images, around dwarf galaxies which have significantly smaller stellar masses than the blue ellipticals in this study. Since tidal debris tends to become fainter when lower mass galaxies are involved, this strengthens the conclusion that, if they were present, tidal features would indeed be visible around our blue ellipticals.} Figures \ref{fig:gals2} and \ref{fig:gals3} show the blue ellipticals which are relaxed and show no tidal perturbations and the three objects that do, respectively. 

Note that, if the star formation is indeed merger-driven, then the chances of detecting tidal features is expected to be \textit{higher} when the galaxy is blue, since the star formation episode has started relatively recently and not much time has elapsed in which the tidal features can fade \citep[e.g.][]{Lofthouse2017}. The negligible incidence of tidal features in our blue ellipticals indicates that the star formation observed in these systems is not driven by mergers and interactions but rather by secular gas accretion. This result is consistent with the findings of \citet{Mahajan2018} who combine optical and HI data from the GAMA \citep{Driver2016} and ALFALFA \citep{Giovanelli2005} surveys to reach a similar conclusion about the star formation activity in such systems. Note, however, that while \citet{Mahajan2018} arrive at this conclusion due to blue ellipticals being rich in HI and residing in low-density environments, we are able to probe this directly through the lack of tidal features, by virtue of the depth of our optical images. 

We conclude this section by considering whether our blue ellipticals could have a tidal origin i.e. whether they could have formed in the tidal debris of mergers between two gas-rich massive galaxies \citep[e.g.][]{Bournaud2006,Ploeckinger2018}. The images of our blue ellipticals (Figure \ref{fig:gals2}) demonstrate that none of these galaxies exhibit tidal bridges with nearby galaxies, consistent with the fact that these objects reside further away from the massive galaxies that define the nodes and filaments of the cosmic web (see Section \ref{sec:dens_maps} below). It is also worth noting that the fraction of low mass galaxies that have a tidal origin is likely to be quite low \citep[$\sim$6 per cent, see e.g.][]{Kaviraj2012}. Thus, it is unlikely that our blue ellipticals formed in the tidal debris of mergers between massive galaxies.


\subsection{Blue ellipticals reside further away from nodes and large-scale filaments}
\label{sec:dens_maps}

Recall that Figure \ref{fig:EXMPslice} shows the density map and skeleton calculated using \texttt{DisPerSE} in the redshift range $0.22<z<0.24$, with galaxies of various morphologies identified using our UML algorithm overplotted (other redshift slices are presented in Appendix \ref{app:dens_maps}). Visual inspection of these density maps suggests that, unlike their normal (red) counterparts which tend to populate dense regions (e.g. those around nodes), the blue elliptical population appears to reside preferentially in low-density regions, further away from nodes and large-scale filaments than e.g. normal elliptical galaxies.

These results indicate that our blue ellipticals populate low-density regions of the Universe, at larger projected distances from the cosmic web than other morphological types. 
In Figure \ref{fig:densVSfilDist}, we quantify the locations of the various morphological types within the cosmic web, in terms of local density and the 2D projected physical distance from the nearest nodes and large-scale filaments. Note that this analysis uses all density maps (which span the redshift range $0.18<z<0.31$). \textcolor{black}{While the different morphological types exhibit similar distributions in local density, with their medians coinciding within their statistical uncertainties (top panel of Figure \ref{fig:densVSfilDist}), strong differences are found between the different morphological classes in terms of the projected distances from the nearest nodes and large-scale filaments.} We find that blue ellipticals completely avoid the (projected) inner $\sim$0.5 Mpc around nodes. And while the distribution for normal ellipticals peaks at 0 -- 0.3 Mpc from their nearest large-scale filaments, the corresponding range of values for the blue ellipticals is 0.8 -- 1.3 Mpc (bottom panel of Figure \ref{fig:densVSfilDist}).

Even though such low density regions are relatively poor in gas \citep{Krumm1984,Burchett2020,Li2022}, they are likely to offer the best conditions for star formation in such low-mass objects to be driven via gas accretion. This is because these objects will dominate their immediate environment and will not have to compete with the larger gravitational potential wells of more massive galaxies \citep[e.g.][]{Kreckel2012,Beygu2013,Das2014,Das2015,Florez2021,Jian2022}, which are found closer to large-scale filaments and are more likely to preferentially accrete the gas available in these regions \citep[e.g.][]{Song2021}. While we do not have information about the past assembly history of our blue ellipticals, it is interesting to note that direct gas accretion could also impart their elliptical morphology. This can happen, for example, if the object in question is fed by multiple streams from the cosmic web. The net angular momentum of the gas sourced from multiple streams can be low because the angular momenta imparted by individual streams can largely cancel out \citep[e.g.][]{Odekon2018,Song2021}, resulting in new stars with a mass profile resembling that found in ellipticals (rather than a disc).


\subsection{Blue ellipticals as progenitors of the general low mass elliptical population} 
\label{sec:growth}

We complete our study by exploring the connection between our blue ellipticals and the general low-mass elliptical galaxy population. We consider the 25$^{\rm th}$ and 75$^{\rm th}$ percentile values of the distributions of projected distances from nearest filaments and nodes of the blue ellipticals, which translates to distances between 0.7 and 1.6 Mpc. The ratio of blue to normal ellipticals that lie both within these distances and within the stellar mass range of the blue ellipticals (10$^8$ < M$_*$ < 10$^{9.5}$ M$_\odot$), is around a factor of 2. 

\begin{figure}
    \centering
    \includegraphics[width=0.45\textwidth]{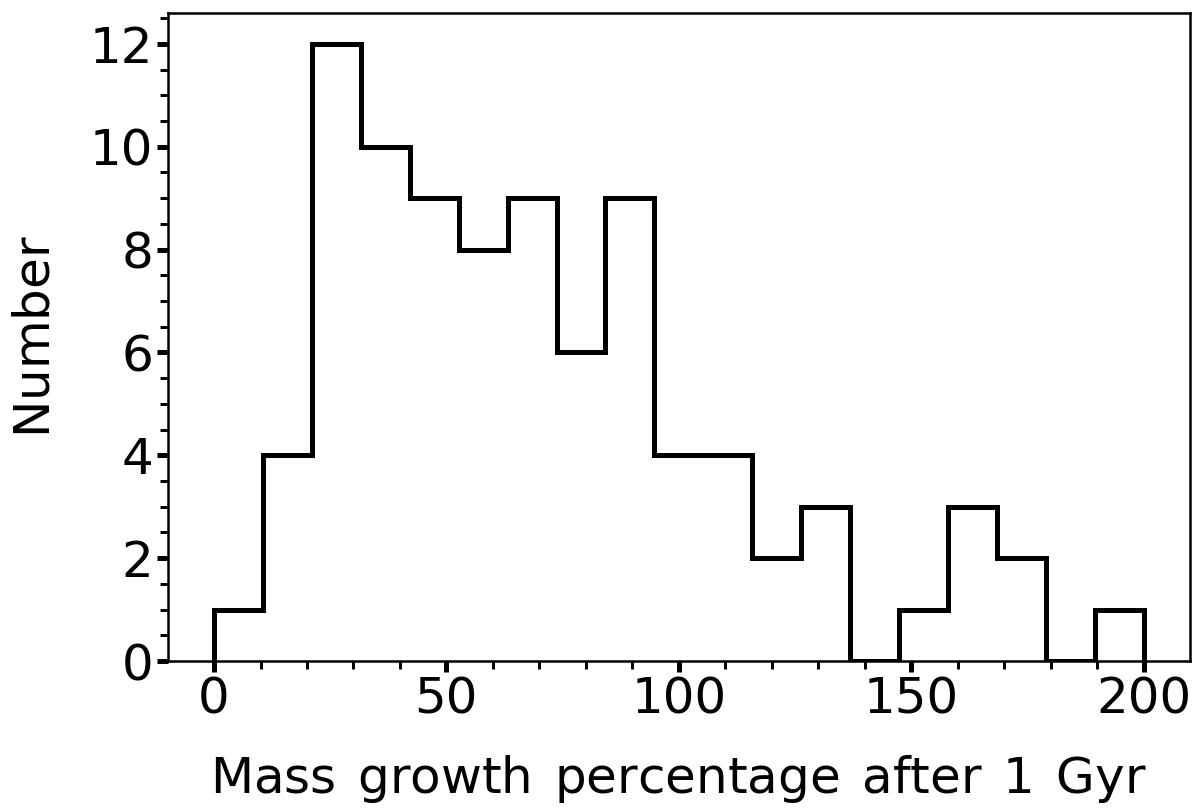}
    \caption{Mass growth percentage in our blue ellipticals, relative to their current stellar mass, if their current measured SFRs remain constant for 1 Gyr.}
    \label{fig:growthRate3}
\end{figure}

\citet{Kaviraj2007} show that galaxies in the blue cloud, in which star formation has been completely shut off, migrate out of the blue cloud over timescales of $\sim$1.5 Gyrs. This is consistent with the findings of \citet{Schawinski2007}, who suggest that blue ellipticals likely remain in the blue cloud for at least around a Gyr, after which they migrate towards the red sequence, either as a result of natural or possibly feedback-driven gas exhaustion. In Figure \ref{fig:growthRate3} we show the stellar mass growth that can be expected in our blue ellipticals if their current measured SFR remains constant over 1 Gyr. The distribution of mass growth values indicates that some blue ellipticals can increase their stellar mass by 100 per cent or more i.e. they are capable of more than doubling their stellar mass over a timescale of around a Gyr.

The number of blue ellipticals is sufficient for these galaxies to feed the population of normal ellipticals which reside in similar environments and have similar (or somewhat higher) stellar masses. In addition, in the context of hierarchical structure formation, galaxies will tend to migrate towards higher-density environments, so that ellipticals which currently reside in low-density environments will progressively populate higher-density environments over cosmic time. In that sense, our blue ellipticals may also feed the elliptical population in denser environments within the cosmic web. 

\textcolor{black}{It is worth noting here that theoretical studies of the formation of elliptical galaxies, using cosmological simulations \citep[e.g.][]{Kobayashi2004,Kobayashi2005,Feldmann2010,Merlin2012,Rodriguez-Gomez2016}, appear aligned with our observational results.} \textcolor{black}{These studies show that, while most massive (M$_\star$ $>$ 10$^{10.5}$ M$_\odot$) ellipticals  undergo (major) mergers in their formation histories, their lower mass counterparts have relatively relaxed, merger-free assembly histories}. Taken together with our results, this suggests that accretion-driven stellar assembly, without recourse to mergers, could be a significant channel for stellar mass assembly in the low mass elliptical galaxy population. 

\section{Summary}
\label{sec:summary}

The evolution of elliptical galaxies is a key topic in observational cosmology. Within the standard hierarchical paradigm, the formation of these systems, in the high-mass regime, is strongly linked to galaxy mergers and interactions. However, the low mass domain remains less well explored and empirical studies that probe the physical processes that drive the evolution of low-mass elliptical galaxies are highly desirable. In particular, studying populations of ellipticals when they are blue (and therefore actively building significant fractions of their stellar mass) can offer important insights into their formation. 

Here, we have studied a sample of 108 blue ellipticals in the low mass regime (with a median mass of $\sim$10$^{8.7}$ M$_{\odot}$) at $z<0.3$. The novelties of this study include (1) making use of extremely deep images from the HSC-SSP (which are around 6 magnitudes deeper than standard-depth imaging from the SDSS), (2) exploiting 40 filter photometry that yields accurate physical parameters for galaxies (e.g. redshifts, stellar masses, SFRs and rest-frame colours) via the COSMOS2020 catalog and (3) quantitatively investigating the local density and locations in the cosmic web of our blue elliptical sample, using the DisPerSE structure finding algorithm. Our main results are as follows: 

\begin{itemize}
    \item Less than 3 per cent of our blue ellipticals show signs of tidal features (indicative of recent mergers and interactions), which are expected to be readily visible at the depth of the images used in this study. The incidence of tidal features in the blue ellipticals is also a factor of 2 lower than in the general galaxy population, strongly implying that the star formation activity is likely to be driven, not by mergers or interactions, but by secular gas accretion.      
    
    \item The blue ellipticals typically inhabit low-density environments and reside further away from nodes and filaments of the cosmic web compared to other morphological types. \textcolor{black}{Such environments are likely to facilitate sustained and efficient gas accretion on to these systems because they are able to dominate their local environment, as opposed to higher density regions where nearby massive galaxies will preferentially accrete the available gas by virtue of their deeper potential wells.} 
    
    \item The number ratios of blue and normal (non-blue) ellipticals that have similar stellar masses and reside in similar environments is around a factor of 2. In addition, given their high SFRs, our blue ellipticals have the potential to significantly increase their stellar mass over timescales of $\sim$ 1 Gyr. Thus, these blue ellipticals are likely progenitors of typical ellipticals at low (and slightly higher) stellar masses.
    
    \item Taken together, our results suggest that stellar mass growth driven by secular gas accretion plays a significant (and possibly dominant) role in the stellar mass build up of elliptical galaxies in the low mass regime.
\end{itemize}


\section*{Acknowledgements}
We thank the referee for numerous constructive comments which improved the original manuscript. We also thank Chiaki Kobayashi for many interesting discussions. IL acknowledges a PhD studentship funded by the Centre for Astrophysics Research at the University of Hertfordshire. SK acknowledges support from the STFC [ST/S00615X/1] and a Senior Research Fellowship from Worcester College Oxford. CL acknowledges support from the Programme National Cosmology and Galaxies (PNCG) of CNRS/INSU with INP and IN2P3, co-funded by CEA and CNES. For the purpose of open access, the author has applied a Creative Commons Attribution (CC BY) licence to any Author Accepted Manuscript version arising from this submission.  


\section*{Data Availability}
The physical parameters of galaxies used in this study are publicly available as part of the COSMOS2020 catalog \citep{Weaver2022}. Galaxy morphological classifications using the unsupervised machine learning code used in this study will be released in a forthcoming paper (Lazar et al. in prep). Readers interested in using these classifications before publication are welcome to contact the authors. 


\bibliographystyle{mnras}
\bibliography{references}

\begin{thebibliography}{}
\makeatletter
\relax
\def\mn@urlcharsother{\let\do\@makeother \do\$\do\&\do\#\do\^\do\_\do\%\do\~}
\def\mn@doi{\begingroup\mn@urlcharsother \@ifnextchar [ {\mn@doi@}
  {\mn@doi@[]}}
\def\mn@doi@[#1]#2{\def\@tempa{#1}\ifx\@tempa\@empty \href
  {http://dx.doi.org/#2} {doi:#2}\else \href {http://dx.doi.org/#2} {#1}\fi
  \endgroup}
\def\mn@eprint#1#2{\mn@eprint@#1:#2::\@nil}
\def\mn@eprint@arXiv#1{\href {http://arxiv.org/abs/#1} {{\tt arXiv:#1}}}
\def\mn@eprint@dblp#1{\href {http://dblp.uni-trier.de/rec/bibtex/#1.xml}
  {dblp:#1}}
\def\mn@eprint@#1:#2:#3:#4\@nil{\def\@tempa {#1}\def\@tempb {#2}\def\@tempc
  {#3}\ifx \@tempc \@empty \let \@tempc \@tempb \let \@tempb \@tempa \fi \ifx
  \@tempb \@empty \def\@tempb {arXiv}\fi \@ifundefined
  {mn@eprint@\@tempb}{\@tempb:\@tempc}{\expandafter \expandafter \csname
  mn@eprint@\@tempb\endcsname \expandafter{\@tempc}}}

\bibitem[\protect\citeauthoryear{{Aihara} et~al.,}{{Aihara}
  et~al.}{2022}]{Aihara2022}
{Aihara} H.,  et~al., 2022, \mn@doi [\pasj] {10.1093/pasj/psab122}, \href
  {https://ui.adsabs.harvard.edu/abs/2022PASJ...74..247A} {74, 247}

\bibitem[\protect\citeauthoryear{{Bailer-Jones}, {Fouesneau}  \&
  {Andrae}}{{Bailer-Jones} et~al.}{2019}]{Bailer2019}
{Bailer-Jones} C. A.~L.,  {Fouesneau} M.,   {Andrae} R.,  2019, \mn@doi
  [\mnras] {10.1093/mnras/stz2947}, \href
  {https://ui.adsabs.harvard.edu/abs/2019MNRAS.490.5615B} {490, 5615}

\bibitem[\protect\citeauthoryear{{Baldry}}{{Baldry}}{2007}]{Baldry2007}
{Baldry} I.~K.,  2007, in {Metcalfe} N.,  {Shanks} T.,  eds,  Astronomical
  Society of the Pacific Conference Series Vol. 379, Cosmic Frontiers. p.~276

\bibitem[\protect\citeauthoryear{{Bamford} et~al.,}{{Bamford}
  et~al.}{2009}]{Bamford2009}
{Bamford} S.~P.,  et~al., 2009, \mn@doi [\mnras]
  {10.1111/j.1365-2966.2008.14252.x}, \href
  {https://ui.adsabs.harvard.edu/abs/2009MNRAS.393.1324B} {393, 1324}

\bibitem[\protect\citeauthoryear{{Bender}}{{Bender}}{1988}]{Bender1988}
{Bender} R.,  1988, \aap, \href
  {https://ui.adsabs.harvard.edu/abs/1988A&A...202L...5B} {202, L5}

\bibitem[\protect\citeauthoryear{{Beygu}, {Kreckel}, {van de Weygaert}, {van
  der Hulst}  \& {van Gorkom}}{{Beygu} et~al.}{2013}]{Beygu2013}
{Beygu} B.,  {Kreckel} K.,  {van de Weygaert} R.,  {van der Hulst} J.~M.,
  {van Gorkom} J.~H.,  2013, \mn@doi [\aj] {10.1088/0004-6256/145/5/120}, \href
  {https://ui.adsabs.harvard.edu/abs/2013AJ....145..120B} {145, 120}

\bibitem[\protect\citeauthoryear{{B{\'\i}lek}, {Fensch}, {Ebrov{\'a}},
  {Nagesh}, {Famaey}, {Duc}  \& {Kroupa}}{{B{\'\i}lek}
  et~al.}{2022}]{Bilek2022}
{B{\'\i}lek} M.,  {Fensch} J.,  {Ebrov{\'a}} I.,  {Nagesh} S.~T.,  {Famaey} B.,
   {Duc} P.-A.,   {Kroupa} P.,  2022, \mn@doi [\aap]
  {10.1051/0004-6361/202141709}, \href
  {https://ui.adsabs.harvard.edu/abs/2022A&A...660A..28B} {660, A28}

\bibitem[\protect\citeauthoryear{{Blanton} et~al.,}{{Blanton}
  et~al.}{2003}]{Blanton2003}
{Blanton} M.~R.,  et~al., 2003, \mn@doi [\apj] {10.1086/375528}, \href
  {https://ui.adsabs.harvard.edu/abs/2003ApJ...594..186B} {594, 186}

\bibitem[\protect\citeauthoryear{{Blumenthal}, {Faber}, {Primack}  \&
  {Rees}}{{Blumenthal} et~al.}{1984}]{Blumenthal1984}
{Blumenthal} G.~R.,  {Faber} S.~M.,  {Primack} J.~R.,   {Rees} M.~J.,  1984,
  \mn@doi [\nat] {10.1038/311517a0}, \href
  {https://ui.adsabs.harvard.edu/abs/1984Natur.311..517B} {311, 517}

\bibitem[\protect\citeauthoryear{{Bois} et~al.,}{{Bois}
  et~al.}{2011}]{Bois2011}
{Bois} M.,  et~al., 2011, \mn@doi [\mnras] {10.1111/j.1365-2966.2011.19113.x},
  \href {https://ui.adsabs.harvard.edu/abs/2011MNRAS.416.1654B} {416, 1654}

\bibitem[\protect\citeauthoryear{{Bournaud} \& {Duc}}{{Bournaud} \&
  {Duc}}{2006}]{Bournaud2006}
{Bournaud} F.,  {Duc} P.~A.,  2006, \mn@doi [\aap]
  {10.1051/0004-6361:20065248}, \href
  {https://ui.adsabs.harvard.edu/abs/2006A&A...456..481B} {456, 481}

\bibitem[\protect\citeauthoryear{{Boylan-Kolchin}, {Ma}  \&
  {Quataert}}{{Boylan-Kolchin} et~al.}{2006}]{Boylan-Kolchin2006}
{Boylan-Kolchin} M.,  {Ma} C.-P.,   {Quataert} E.,  2006, \mn@doi [\mnras]
  {10.1111/j.1365-2966.2006.10379.x}, \href
  {https://ui.adsabs.harvard.edu/abs/2006MNRAS.369.1081B} {369, 1081}

\bibitem[\protect\citeauthoryear{{Brough}, {Forbes}, {Kilborn}  \&
  {Couch}}{{Brough} et~al.}{2006}]{Brough2006}
{Brough} S.,  {Forbes} D.~A.,  {Kilborn} V.~A.,   {Couch} W.,  2006, \mn@doi
  [\mnras] {10.1111/j.1365-2966.2006.10542.x}, \href
  {https://ui.adsabs.harvard.edu/abs/2006MNRAS.370.1223B} {370, 1223}

\bibitem[\protect\citeauthoryear{{Burchett}, {Elek}, {Tejos}, {Prochaska},
  {Tripp}, {Bordoloi}  \& {Forbes}}{{Burchett} et~al.}{2020}]{Burchett2020}
{Burchett} J.~N.,  {Elek} O.,  {Tejos} N.,  {Prochaska} J.~X.,  {Tripp} T.~M.,
  {Bordoloi} R.,   {Forbes} A.~G.,  2020, \mn@doi [\apjl]
  {10.3847/2041-8213/ab700c}, \href
  {https://ui.adsabs.harvard.edu/abs/2020ApJ...891L..35B} {891, L35}

\bibitem[\protect\citeauthoryear{{Cameron}}{{Cameron}}{2011}]{Cameron2011}
{Cameron} E.,  2011, \mn@doi [\pasa] {10.1071/AS10046}, \href
  {https://ui.adsabs.harvard.edu/abs/2011PASA...28..128C} {28, 128}

\bibitem[\protect\citeauthoryear{{Cappellari} et~al.,}{{Cappellari}
  et~al.}{2011}]{Cappellari2011}
{Cappellari} M.,  et~al., 2011, \mn@doi [\mnras]
  {10.1111/j.1365-2966.2011.18600.x}, \href
  {http://adsabs.harvard.edu/abs/2011MNRAS.416.1680C} {416, 1680}

\bibitem[\protect\citeauthoryear{{Chiosi} \& {Carraro}}{{Chiosi} \&
  {Carraro}}{2002}]{Chiosi2002}
{Chiosi} C.,  {Carraro} G.,  2002, \mn@doi [\mnras]
  {10.1046/j.1365-8711.2002.05590.x}, \href
  {https://ui.adsabs.harvard.edu/abs/2002MNRAS.335..335C} {335, 335}

\bibitem[\protect\citeauthoryear{{Chung}, {Rey}, {Sung}, {Kim}, {Lee}  \&
  {Lee}}{{Chung} et~al.}{2019}]{Chung2019}
{Chung} J.,  {Rey} S.-C.,  {Sung} E.-C.,  {Kim} S.,  {Lee} Y.,   {Lee} W.,
  2019, \mn@doi [\apj] {10.3847/1538-4357/ab25e8}, \href
  {https://ui.adsabs.harvard.edu/abs/2019ApJ...879...97C} {879, 97}

\bibitem[\protect\citeauthoryear{{Conselice}}{{Conselice}}{2003}]{Conselice2003}
{Conselice} C.~J.,  2003, \mn@doi [\apjs] {10.1086/375001}, \href
  {http://adsabs.harvard.edu/abs/2003ApJS..147....1C} {147, 1}

\bibitem[\protect\citeauthoryear{{Cooper} et~al.,}{{Cooper}
  et~al.}{2011}]{Cooper2011}
{Cooper} A.~P.,  et~al., 2011, \mn@doi [\apjl] {10.1088/2041-8205/743/1/L21},
  \href {https://ui.adsabs.harvard.edu/abs/2011ApJ...743L..21C} {743, L21}

\bibitem[\protect\citeauthoryear{{Crone Odekon}, {Hallenbeck}, {Haynes},
  {Koopmann}, {Phi}  \& {Wolfe}}{{Crone Odekon} et~al.}{2018}]{Odekon2018}
{Crone Odekon} M.,  {Hallenbeck} G.,  {Haynes} M.~P.,  {Koopmann} R.~A.,  {Phi}
  A.,   {Wolfe} P.-F.,  2018, \mn@doi [\apj] {10.3847/1538-4357/aaa1e8}, \href
  {https://ui.adsabs.harvard.edu/abs/2018ApJ...852..142C} {852, 142}

\bibitem[\protect\citeauthoryear{{Darg} et~al.,}{{Darg}
  et~al.}{2010}]{Darg2010}
{Darg} D.~W.,  et~al., 2010, \mn@doi [\mnras]
  {10.1111/j.1365-2966.2009.15786.x}, \href
  {https://ui.adsabs.harvard.edu/abs/2010MNRAS.401.1552D} {401, 1552}

\bibitem[\protect\citeauthoryear{{Das}, {Iono}, {Saito}  \&
  {Subramanian}}{{Das} et~al.}{2014}]{Das2014}
{Das} M.,  {Iono} D.,  {Saito} T.,   {Subramanian} S.,  2014, in Astronomical
  Society of India Conference Series. pp 299--301

\bibitem[\protect\citeauthoryear{{Das}, {Saito}, {Iono}, {Honey}  \&
  {Ramya}}{{Das} et~al.}{2015}]{Das2015}
{Das} M.,  {Saito} T.,  {Iono} D.,  {Honey} M.,   {Ramya} S.,  2015, \mn@doi
  [\apj] {10.1088/0004-637X/815/1/40}, \href
  {https://ui.adsabs.harvard.edu/abs/2015ApJ...815...40D} {815, 40}

\bibitem[\protect\citeauthoryear{{Davis} et~al.,}{{Davis}
  et~al.}{2022}]{Davis2022}
{Davis} F.,  et~al., 2022, \mn@doi [\mnras] {10.1093/mnras/stac068}, \href
  {https://ui.adsabs.harvard.edu/abs/2022MNRAS.511.4109D} {511, 4109}

\bibitem[\protect\citeauthoryear{{De Propris}, {Bremer}  \& {Phillipps}}{{De
  Propris} et~al.}{2016}]{Propris2016}
{De Propris} R.,  {Bremer} M.~N.,   {Phillipps} S.,  2016, \mn@doi [\mnras]
  {10.1093/mnras/stw1521}, \href
  {https://ui.adsabs.harvard.edu/abs/2016MNRAS.461.4517D} {461, 4517}

\bibitem[\protect\citeauthoryear{{Deeley} et~al.,}{{Deeley}
  et~al.}{2017}]{Deeley2017}
{Deeley} S.,  et~al., 2017, \mn@doi [\mnras] {10.1093/mnras/stx441}, \href
  {http://adsabs.harvard.edu/abs/2017MNRAS.467.3934D} {467, 3934}

\bibitem[\protect\citeauthoryear{{Dekel} \& {Birnboim}}{{Dekel} \&
  {Birnboim}}{2006}]{Dekel2006}
{Dekel} A.,  {Birnboim} Y.,  2006, \mn@doi [\mnras]
  {10.1111/j.1365-2966.2006.10145.x}, \href
  {https://ui.adsabs.harvard.edu/abs/2006MNRAS.368....2D} {368, 2}

\bibitem[\protect\citeauthoryear{{Dressler}}{{Dressler}}{1980}]{Dressler1980}
{Dressler} A.,  1980, \mn@doi [\apj] {10.1086/157753}, \href
  {https://ui.adsabs.harvard.edu/abs/1980ApJ...236..351D} {236, 351}

\bibitem[\protect\citeauthoryear{{Driver} et~al.,}{{Driver}
  et~al.}{2016}]{Driver2016}
{Driver} S.~P.,  et~al., 2016, \mn@doi [\mnras] {10.1093/mnras/stv2505}, \href
  {https://ui.adsabs.harvard.edu/abs/2016MNRAS.455.3911D} {455, 3911}

\bibitem[\protect\citeauthoryear{{Dubois} et~al.,}{{Dubois}
  et~al.}{2014}]{Dubois2014}
{Dubois} Y.,  et~al., 2014, \mn@doi [\mnras] {10.1093/mnras/stu1227}, \href
  {http://adsabs.harvard.edu/abs/2014MNRAS.444.1453D} {444, 1453}

\bibitem[\protect\citeauthoryear{{Dubois} et~al.,}{{Dubois}
  et~al.}{2021}]{Dubois2021}
{Dubois} Y.,  et~al., 2021, \mn@doi [\aap] {10.1051/0004-6361/202039429}, \href
  {https://ui.adsabs.harvard.edu/abs/2021A&A...651A.109D} {651, A109}

\bibitem[\protect\citeauthoryear{{Erwin}}{{Erwin}}{2015}]{Erwin2015}
{Erwin} P.,  2015, \mn@doi [\apj] {10.1088/0004-637X/799/2/226}, \href
  {https://ui.adsabs.harvard.edu/abs/2015ApJ...799..226E} {799, 226}

\bibitem[\protect\citeauthoryear{{Faber} et~al.,}{{Faber}
  et~al.}{2007}]{Faber2007}
{Faber} S.~M.,  et~al., 2007, \mn@doi [\apj] {10.1086/519294}, \href
  {https://ui.adsabs.harvard.edu/abs/2007ApJ...665..265F} {665, 265}

\bibitem[\protect\citeauthoryear{{Feldmann}, {Carollo}, {Mayer}, {Renzini},
  {Lake}, {Quinn}, {Stinson}  \& {Yepes}}{{Feldmann}
  et~al.}{2010}]{Feldmann2010}
{Feldmann} R.,  {Carollo} C.~M.,  {Mayer} L.,  {Renzini} A.,  {Lake} G.,
  {Quinn} T.,  {Stinson} G.~S.,   {Yepes} G.,  2010, \mn@doi [\apj]
  {10.1088/0004-637X/709/1/218}, \href
  {https://ui.adsabs.harvard.edu/abs/2010ApJ...709..218F} {709, 218}

\bibitem[\protect\citeauthoryear{{Fensch} et~al.,}{{Fensch}
  et~al.}{2017}]{Fensch2017}
{Fensch} J.,  et~al., 2017, \mn@doi [\mnras] {10.1093/mnras/stw2920}, \href
  {http://adsabs.harvard.edu/abs/2017MNRAS.465.1934F} {465, 1934}

\bibitem[\protect\citeauthoryear{{Ferreras}, {Lisker}, {Carollo}, {Lilly}  \&
  {Mobasher}}{{Ferreras} et~al.}{2005}]{Ferreras2005}
{Ferreras} I.,  {Lisker} T.,  {Carollo} C.~M.,  {Lilly} S.~J.,   {Mobasher} B.,
   2005, \mn@doi [\apj] {10.1086/497292}, \href
  {https://ui.adsabs.harvard.edu/abs/2005ApJ...635..243F} {635, 243}

\bibitem[\protect\citeauthoryear{{Florez} et~al.,}{{Florez}
  et~al.}{2021}]{Florez2021}
{Florez} J.,  et~al., 2021, \mn@doi [\apj] {10.3847/1538-4357/abca9f}, \href
  {https://ui.adsabs.harvard.edu/abs/2021ApJ...906...97F} {906, 97}

\bibitem[\protect\citeauthoryear{{Forbes}, {Franx}  \& {Illingworth}}{{Forbes}
  et~al.}{1995}]{Forbes1995}
{Forbes} D.~A.,  {Franx} M.,   {Illingworth} G.~D.,  1995, \mn@doi [\aj]
  {10.1086/117425}, \href
  {https://ui.adsabs.harvard.edu/abs/1995AJ....109.1988F} {109, 1988}

\bibitem[\protect\citeauthoryear{Fritzke}{Fritzke}{1995}]{Fritzke1995}
Fritzke B.,  1995, in Advances in neural information processing systems. pp
  625--632

\bibitem[\protect\citeauthoryear{{Fukugita}, {Nakamura}, {Turner}, {Helmboldt}
  \& {Nichol}}{{Fukugita} et~al.}{2004}]{Fukugita2004}
{Fukugita} M.,  {Nakamura} O.,  {Turner} E.~L.,  {Helmboldt} J.,   {Nichol}
  R.~C.,  2004, \mn@doi [\apjl] {10.1086/382151}, \href
  {https://ui.adsabs.harvard.edu/abs/2004ApJ...601L.127F} {601, L127}

\bibitem[\protect\citeauthoryear{{George}}{{George}}{2017}]{George2017}
{George} K.,  2017, \mn@doi [\aap] {10.1051/0004-6361/201629667}, \href
  {https://ui.adsabs.harvard.edu/abs/2017A&A...598A..45G} {598, A45}

\bibitem[\protect\citeauthoryear{{Giovanelli} et~al.,}{{Giovanelli}
  et~al.}{2005}]{Giovanelli2005}
{Giovanelli} R.,  et~al., 2005, \mn@doi [\aj] {10.1086/497431}, \href
  {https://ui.adsabs.harvard.edu/abs/2005AJ....130.2598G} {130, 2598}

\bibitem[\protect\citeauthoryear{{Gu{\'e}rou} et~al.,}{{Gu{\'e}rou}
  et~al.}{2015}]{Guerou2015}
{Gu{\'e}rou} A.,  et~al., 2015, \mn@doi [\apj] {10.1088/0004-637X/804/1/70},
  \href {https://ui.adsabs.harvard.edu/abs/2015ApJ...804...70G} {804, 70}

\bibitem[\protect\citeauthoryear{{Hocking}, {Geach}, {Sun}  \&
  {Davey}}{{Hocking} et~al.}{2018}]{Hocking2018}
{Hocking} A.,  {Geach} J.~E.,  {Sun} Y.,   {Davey} N.,  2018, \mn@doi [\mnras]
  {10.1093/mnras/stx2351}, \href
  {http://adsabs.harvard.edu/abs/2018MNRAS.473.1108H} {473, 1108}

\bibitem[\protect\citeauthoryear{{Hopkins} et~al.,}{{Hopkins}
  et~al.}{2010}]{Hopkins2010}
{Hopkins} P.~F.,  et~al., 2010, \mn@doi [\apj] {10.1088/0004-637X/715/1/202},
  \href {https://ui.adsabs.harvard.edu/abs/2010ApJ...715..202H} {715, 202}

\bibitem[\protect\citeauthoryear{{Ilbert} et~al.,}{{Ilbert}
  et~al.}{2013}]{Ilbert2013}
{Ilbert} O.,  et~al., 2013, \mn@doi [\aap] {10.1051/0004-6361/201321100}, \href
  {https://ui.adsabs.harvard.edu/abs/2013A&A...556A..55I} {556, A55}

\bibitem[\protect\citeauthoryear{{Jackson} et~al.,}{{Jackson}
  et~al.}{2021}]{Jackson2021}
{Jackson} R.~A.,  et~al., 2021, \mn@doi [\mnras] {10.1093/mnras/stab077}, \href
  {https://ui.adsabs.harvard.edu/abs/2021MNRAS.502.4262J} {502, 4262}

\bibitem[\protect\citeauthoryear{{Jedrzejewski} \& {Schechter}}{{Jedrzejewski}
  \& {Schechter}}{1988}]{Jedrzejewski1988}
{Jedrzejewski} R.,  {Schechter} P.~L.,  1988, \mn@doi [\apjl] {10.1086/185211},
  \href {https://ui.adsabs.harvard.edu/abs/1988ApJ...330L..87J} {330, L87}

\bibitem[\protect\citeauthoryear{{Jian} et~al.,}{{Jian}
  et~al.}{2022}]{Jian2022}
{Jian} H.-Y.,  et~al., 2022, \mn@doi [\apj] {10.3847/1538-4357/ac448b}, \href
  {https://ui.adsabs.harvard.edu/abs/2022ApJ...926..115J} {926, 115}

\bibitem[\protect\citeauthoryear{Johnson}{Johnson}{1967}]{Johnson1967}
Johnson S.~C.,  1967, Psychometrika, 32, 241

\bibitem[\protect\citeauthoryear{{Johnston}, {Zhao}, {Spergel}  \&
  {Hernquist}}{{Johnston} et~al.}{1999}]{Johnston1999}
{Johnston} K.~V.,  {Zhao} H.,  {Spergel} D.~N.,   {Hernquist} L.,  1999,
  \mn@doi [\apjl] {10.1086/311876}, \href
  {https://ui.adsabs.harvard.edu/abs/1999ApJ...512L.109J} {512, L109}

\bibitem[\protect\citeauthoryear{{Kannappan}, {Guie}  \& {Baker}}{{Kannappan}
  et~al.}{2009}]{Kannappan2009}
{Kannappan} S.~J.,  {Guie} J.~M.,   {Baker} A.~J.,  2009, \mn@doi [\aj]
  {10.1088/0004-6256/138/2/579}, \href
  {https://ui.adsabs.harvard.edu/abs/2009AJ....138..579K} {138, 579}

\bibitem[\protect\citeauthoryear{{Kaviraj}}{{Kaviraj}}{2014}]{Kaviraj2014b}
{Kaviraj} S.,  2014, \mn@doi [\mnras] {10.1093/mnras/stu338}, \href
  {http://adsabs.harvard.edu/abs/2014MNRAS.440.2944K} {440, 2944}

\bibitem[\protect\citeauthoryear{{Kaviraj} et~al.,}{{Kaviraj}
  et~al.}{2007}]{Kaviraj2007}
{Kaviraj} S.,  et~al., 2007, \mn@doi [\apjs] {10.1086/516633}, \href
  {https://ui.adsabs.harvard.edu/abs/2007ApJS..173..619K} {173, 619}

\bibitem[\protect\citeauthoryear{{Kaviraj} et~al.,}{{Kaviraj}
  et~al.}{2008}]{Kaviraj2008}
{Kaviraj} S.,  et~al., 2008, \mn@doi [\mnras]
  {10.1111/j.1365-2966.2008.13392.x}, \href
  {https://ui.adsabs.harvard.edu/abs/2008MNRAS.388...67K} {388, 67}

\bibitem[\protect\citeauthoryear{{Kaviraj}, {Tan}, {Ellis}  \&
  {Silk}}{{Kaviraj} et~al.}{2011}]{Kaviraj2011}
{Kaviraj} S.,  {Tan} K.-M.,  {Ellis} R.~S.,   {Silk} J.,  2011, \mn@doi
  [\mnras] {10.1111/j.1365-2966.2010.17754.x}, \href
  {http://adsabs.harvard.edu/abs/2011MNRAS.411.2148K} {411, 2148}

\bibitem[\protect\citeauthoryear{{Kaviraj}, {Darg}, {Lintott}, {Schawinski}  \&
  {Silk}}{{Kaviraj} et~al.}{2012}]{Kaviraj2012}
{Kaviraj} S.,  {Darg} D.,  {Lintott} C.,  {Schawinski} K.,   {Silk} J.,  2012,
  \mn@doi [\mnras] {10.1111/j.1365-2966.2011.19673.x}, \href
  {https://ui.adsabs.harvard.edu/abs/2012MNRAS.419...70K} {419, 70}

\bibitem[\protect\citeauthoryear{{Kaviraj} et~al.,}{{Kaviraj}
  et~al.}{2013}]{Kaviraj2013}
{Kaviraj} S.,  et~al., 2013, \mn@doi [\mnras] {10.1093/mnras/sts031}, \href
  {http://adsabs.harvard.edu/abs/2013MNRAS.428..925K} {428, 925}

\bibitem[\protect\citeauthoryear{{Kaviraj} et~al.,}{{Kaviraj}
  et~al.}{2017}]{Kaviraj2017}
{Kaviraj} S.,  et~al., 2017, \mn@doi [\mnras] {10.1093/mnras/stx126}, \href
  {http://adsabs.harvard.edu/abs/2017MNRAS.467.4739K} {467, 4739}

\bibitem[\protect\citeauthoryear{{Kaviraj}, {Martin}  \& {Silk}}{{Kaviraj}
  et~al.}{2019}]{Kaviraj2019}
{Kaviraj} S.,  {Martin} G.,   {Silk} J.,  2019, \mn@doi [\mnras]
  {10.1093/mnrasl/slz102}, \href
  {https://ui.adsabs.harvard.edu/abs/2019MNRAS.489L..12K} {489, L12}

\bibitem[\protect\citeauthoryear{{Kelvin} et~al.,}{{Kelvin}
  et~al.}{2014}]{Kelvin2014}
{Kelvin} L.~S.,  et~al., 2014, \mn@doi [\mnras] {10.1093/mnras/stu1507}, \href
  {https://ui.adsabs.harvard.edu/abs/2014MNRAS.444.1647K} {444, 1647}

\bibitem[\protect\citeauthoryear{{Kobayashi}}{{Kobayashi}}{2004}]{Kobayashi2004}
{Kobayashi} C.,  2004, \mn@doi [\mnras] {10.1111/j.1365-2966.2004.07258.x},
  \href {https://ui.adsabs.harvard.edu/abs/2004MNRAS.347..740K} {347, 740}

\bibitem[\protect\citeauthoryear{{Kobayashi}}{{Kobayashi}}{2005}]{Kobayashi2005}
{Kobayashi} C.,  2005, \mn@doi [\mnras] {10.1111/j.1365-2966.2005.09248.x},
  \href {https://ui.adsabs.harvard.edu/abs/2005MNRAS.361.1216K} {361, 1216}

\bibitem[\protect\citeauthoryear{{Kolokythas}, {Vaddi}, {O'Sullivan},
  {Loubser}, {Babul}, {Raychaudhury}, {Lagos}  \& {Jarrett}}{{Kolokythas}
  et~al.}{2022}]{Kolokythas2022}
{Kolokythas} K.,  {Vaddi} S.,  {O'Sullivan} E.,  {Loubser} I.,  {Babul} A.,
  {Raychaudhury} S.,  {Lagos} P.,   {Jarrett} T.~H.,  2022, \mn@doi [\mnras]
  {10.1093/mnras/stab3699}, \href
  {https://ui.adsabs.harvard.edu/abs/2022MNRAS.510.4191K} {510, 4191}

\bibitem[\protect\citeauthoryear{{Kormendy}}{{Kormendy}}{1977}]{Kormendy1977}
{Kormendy} J.,  1977, \mn@doi [\apj] {10.1086/155687}, \href
  {https://ui.adsabs.harvard.edu/abs/1977ApJ...218..333K} {218, 333}

\bibitem[\protect\citeauthoryear{{Kreckel}, {Platen}, {Arag{\'o}n-Calvo}, {van
  Gorkom}, {van de Weygaert}, {van der Hulst}  \& {Beygu}}{{Kreckel}
  et~al.}{2012}]{Kreckel2012}
{Kreckel} K.,  {Platen} E.,  {Arag{\'o}n-Calvo} M.~A.,  {van Gorkom} J.~H.,
  {van de Weygaert} R.,  {van der Hulst} J.~M.,   {Beygu} B.,  2012, \mn@doi
  [\aj] {10.1088/0004-6256/144/1/16}, \href
  {https://ui.adsabs.harvard.edu/abs/2012AJ....144...16K} {144, 16}

\bibitem[\protect\citeauthoryear{{Krumm} \& {Brosch}}{{Krumm} \&
  {Brosch}}{1984}]{Krumm1984}
{Krumm} N.,  {Brosch} N.,  1984, \mn@doi [\aj] {10.1086/113647}, \href
  {https://ui.adsabs.harvard.edu/abs/1984AJ.....89.1461K} {89, 1461}

\bibitem[\protect\citeauthoryear{Laigle et~al.,}{Laigle
  et~al.}{2016}]{Laigle2016}
Laigle C.,  et~al., 2016, \mn@doi [The Astrophysical Journal Supplement Series]
  {10.3847/0067-0049/224/2/24}, 224, 24

\bibitem[\protect\citeauthoryear{{Laigle} et~al.,}{{Laigle}
  et~al.}{2018}]{Laigle2018}
{Laigle} C.,  et~al., 2018, \mn@doi [\mnras] {10.1093/mnras/stx3055}, \href
  {https://ui.adsabs.harvard.edu/abs/2018MNRAS.474.5437L} {474, 5437}

\bibitem[\protect\citeauthoryear{{Larson}}{{Larson}}{1974}]{Larson1974}
{Larson} R.~B.,  1974, \mn@doi [\mnras] {10.1093/mnras/166.3.585}, \href
  {https://ui.adsabs.harvard.edu/abs/1974MNRAS.166..585L} {166, 585}

\bibitem[\protect\citeauthoryear{{Li}, {Yee}, {Hsieh}  \& {Gladders}}{{Li}
  et~al.}{2012}]{Li2012}
{Li} I.~H.,  {Yee} H.~K.~C.,  {Hsieh} B.~C.,   {Gladders} M.,  2012, \mn@doi
  [\apj] {10.1088/0004-637X/749/2/150}, \href
  {https://ui.adsabs.harvard.edu/abs/2012ApJ...749..150L} {749, 150}

\bibitem[\protect\citeauthoryear{{Li} et~al.,}{{Li} et~al.}{2022}]{Li2022}
{Li} R.,  et~al., 2022, \mn@doi [\apj] {10.3847/1538-4357/ac8359}, \href
  {https://ui.adsabs.harvard.edu/abs/2022ApJ...936...11L} {936, 11}

\bibitem[\protect\citeauthoryear{{Lofthouse}, {Kaviraj}, {Conselice},
  {Mortlock}  \& {Hartley}}{{Lofthouse} et~al.}{2017}]{Lofthouse2017}
{Lofthouse} E.~K.,  {Kaviraj} S.,  {Conselice} C.~J.,  {Mortlock} A.,
  {Hartley} W.,  2017, \mn@doi [\mnras] {10.1093/mnras/stw2895}, \href
  {http://adsabs.harvard.edu/abs/2017MNRAS.465.2895L} {465, 2895}

\bibitem[\protect\citeauthoryear{{Mahajan} et~al.,}{{Mahajan}
  et~al.}{2018}]{Mahajan2018}
{Mahajan} S.,  et~al., 2018, \mn@doi [\mnras] {10.1093/mnras/stx3202}, \href
  {https://ui.adsabs.harvard.edu/abs/2018MNRAS.475..788M} {475, 788}

\bibitem[\protect\citeauthoryear{{Martin}, {Kaviraj}, {Devriendt}, {Dubois}  \&
  {Pichon}}{{Martin} et~al.}{2018}]{Martin2018a}
{Martin} G.,  {Kaviraj} S.,  {Devriendt} J.~E.~G.,  {Dubois} Y.,   {Pichon} C.,
   2018, \mn@doi [\mnras] {10.1093/mnras/sty1936}, \href
  {http://adsabs.harvard.edu/abs/2018MNRAS.480.2266M} {480, 2266}

\bibitem[\protect\citeauthoryear{{Martin} et~al.,}{{Martin}
  et~al.}{2019}]{Martin2019}
{Martin} G.,  et~al., 2019, \mn@doi [\mnras] {10.1093/mnras/stz356}, \href
  {http://adsabs.harvard.edu/abs/2019MNRAS.485..796M} {485, 796}

\bibitem[\protect\citeauthoryear{{Martin}, {Kaviraj}, {Hocking}, {Read}  \&
  {Geach}}{{Martin} et~al.}{2020}]{Martin2020}
{Martin} G.,  {Kaviraj} S.,  {Hocking} A.,  {Read} S.~C.,   {Geach} J.~E.,
  2020, \mn@doi [\mnras] {10.1093/mnras/stz3006}, \href
  {https://ui.adsabs.harvard.edu/abs/2020MNRAS.491.1408M} {491, 1408}

\bibitem[\protect\citeauthoryear{{Martin} et~al.,}{{Martin}
  et~al.}{2022}]{Martin2022}
{Martin} G.,  et~al., 2022, \mn@doi [\mnras] {10.1093/mnras/stac1003}, \href
  {https://ui.adsabs.harvard.edu/abs/2022MNRAS.513.1459M} {513, 1459}

\bibitem[\protect\citeauthoryear{{Merlin}, {Chiosi}, {Piovan}, {Grassi},
  {Buonomo}  \& {La Barbera}}{{Merlin} et~al.}{2012}]{Merlin2012}
{Merlin} E.,  {Chiosi} C.,  {Piovan} L.,  {Grassi} T.,  {Buonomo} U.,   {La
  Barbera} F.,  2012, \mn@doi [\mnras] {10.1111/j.1365-2966.2012.21965.x},
  \href {https://ui.adsabs.harvard.edu/abs/2012MNRAS.427.1530M} {427, 1530}

\bibitem[\protect\citeauthoryear{{Meyer}, {Lisker}, {Janz}  \&
  {Papaderos}}{{Meyer} et~al.}{2014}]{Meyer2014}
{Meyer} H.~T.,  {Lisker} T.,  {Janz} J.,   {Papaderos} P.,  2014, \mn@doi
  [\aap] {10.1051/0004-6361/201220700}, \href
  {https://ui.adsabs.harvard.edu/abs/2014A&A...562A..49M} {562, A49}

\bibitem[\protect\citeauthoryear{{Mihos} \& {Hernquist}}{{Mihos} \&
  {Hernquist}}{1996}]{Mihos1996}
{Mihos} J.~C.,  {Hernquist} L.,  1996, \mn@doi [\apj] {10.1086/177353}, \href
  {http://adsabs.harvard.edu/abs/1996ApJ...464..641M} {464, 641}

\bibitem[\protect\citeauthoryear{{Miller}, {Roelofs}  \& {Smith}}{{Miller}
  et~al.}{1990}]{Miller1990}
{Miller} R.~H.,  {Roelofs} G.~R.,   {Smith} B.~F.,  1990, in {Sulentic} J.~W.,
  {Keel} W.~C.,   {Telesco} C.~M.,  eds,  NASA Conference Publication Vol.
  3098, NASA Conference Publication. pp 549--554

\bibitem[\protect\citeauthoryear{{Moffett} et~al.,}{{Moffett}
  et~al.}{2019}]{Moffett2019}
{Moffett} A.~J.,  et~al., 2019, \mn@doi [\mnras] {10.1093/mnras/stz2237}, \href
  {https://ui.adsabs.harvard.edu/abs/2019MNRAS.489.2830M} {489, 2830}

\bibitem[\protect\citeauthoryear{{Moore}, {Katz}, {Lake}, {Dressler}  \&
  {Oemler}}{{Moore} et~al.}{1996}]{Moore1996}
{Moore} B.,  {Katz} N.,  {Lake} G.,  {Dressler} A.,   {Oemler} A.,  1996,
  \mn@doi [\nat] {10.1038/379613a0}, \href
  {https://ui.adsabs.harvard.edu/abs/1996Natur.379..613M} {379, 613}

\bibitem[\protect\citeauthoryear{{Omar} \& {Paswan}}{{Omar} \&
  {Paswan}}{2018}]{Omar2018}
{Omar} A.,  {Paswan} A.,  2018, \mn@doi [\mnras] {10.1093/mnras/sty740}, \href
  {https://ui.adsabs.harvard.edu/abs/2018MNRAS.477.3552O} {477, 3552}

\bibitem[\protect\citeauthoryear{{Pannella}, {Hopp}, {Saglia}, {Bender},
  {Drory}, {Salvato}, {Gabasch}  \& {Feulner}}{{Pannella}
  et~al.}{2006}]{Pannella2006}
{Pannella} M.,  {Hopp} U.,  {Saglia} R.~P.,  {Bender} R.,  {Drory} N.,
  {Salvato} M.,  {Gabasch} A.,   {Feulner} G.,  2006, \mn@doi [\apjl]
  {10.1086/501452}, \href
  {https://ui.adsabs.harvard.edu/abs/2006ApJ...639L...1P} {639, L1}

\bibitem[\protect\citeauthoryear{{Park} et~al.,}{{Park}
  et~al.}{2019}]{Park2019}
{Park} M.-J.,  et~al., 2019, \mn@doi [\apj] {10.3847/1538-4357/ab3afe}, \href
  {https://ui.adsabs.harvard.edu/abs/2019ApJ...883...25P} {883, 25}

\bibitem[\protect\citeauthoryear{{Peirani}, {Crockett}, {Geen}, {Khochfar},
  {Kaviraj}  \& {Silk}}{{Peirani} et~al.}{2010}]{Peirani2010}
{Peirani} S.,  {Crockett} R.~M.,  {Geen} S.,  {Khochfar} S.,  {Kaviraj} S.,
  {Silk} J.,  2010, \mn@doi [\mnras] {10.1111/j.1365-2966.2010.16666.x}, \href
  {https://ui.adsabs.harvard.edu/abs/2010MNRAS.405.2327P} {405, 2327}

\bibitem[\protect\citeauthoryear{{Ploeckinger}, {Sharma}, {Schaye}, {Crain},
  {Schaller}  \& {Barber}}{{Ploeckinger} et~al.}{2018}]{Ploeckinger2018}
{Ploeckinger} S.,  {Sharma} K.,  {Schaye} J.,  {Crain} R.~A.,  {Schaller} M.,
  {Barber} C.,  2018, \mn@doi [\mnras] {10.1093/mnras/stx2787}, \href
  {https://ui.adsabs.harvard.edu/abs/2018MNRAS.474..580P} {474, 580}

\bibitem[\protect\citeauthoryear{{Press} \& {Schechter}}{{Press} \&
  {Schechter}}{1974}]{Press1974}
{Press} W.~H.,  {Schechter} P.,  1974, \mn@doi [\apj] {10.1086/152650}, \href
  {https://ui.adsabs.harvard.edu/abs/1974ApJ...187..425P} {187, 425}

\bibitem[\protect\citeauthoryear{{Quilley} \& {de Lapparent}}{{Quilley} \& {de
  Lapparent}}{2022}]{Quilley2022}
{Quilley} L.,  {de Lapparent} V.,  2022, arXiv e-prints, \href
  {https://ui.adsabs.harvard.edu/abs/2022arXiv220604707Q} {p. arXiv:2206.04707}

\bibitem[\protect\citeauthoryear{{Quinn}}{{Quinn}}{1984}]{Quinn1984}
{Quinn} P.~J.,  1984, \mn@doi [\apj] {10.1086/161924}, \href
  {https://ui.adsabs.harvard.edu/abs/1984ApJ...279..596Q} {279, 596}

\bibitem[\protect\citeauthoryear{{Rodriguez-Gomez} et~al.,}{{Rodriguez-Gomez}
  et~al.}{2016}]{Rodriguez-Gomez2016}
{Rodriguez-Gomez} V.,  et~al., 2016, \mn@doi [\mnras] {10.1093/mnras/stw456},
  \href {https://ui.adsabs.harvard.edu/abs/2016MNRAS.458.2371R} {458, 2371}

\bibitem[\protect\citeauthoryear{{Schaap} \& {van de Weygaert}}{{Schaap} \&
  {van de Weygaert}}{2000}]{Schaap2000}
{Schaap} W.~E.,  {van de Weygaert} R.,  2000, \aap, \href
  {https://ui.adsabs.harvard.edu/abs/2000A&A...363L..29S} {363, L29}

\bibitem[\protect\citeauthoryear{{Schawinski}, {Thomas}, {Sarzi}, {Maraston},
  {Kaviraj}, {Joo}, {Yi}  \& {Silk}}{{Schawinski}
  et~al.}{2007}]{Schawinski2007}
{Schawinski} K.,  {Thomas} D.,  {Sarzi} M.,  {Maraston} C.,  {Kaviraj} S.,
  {Joo} S.-J.,  {Yi} S.~K.,   {Silk} J.,  2007, \mn@doi [\mnras]
  {10.1111/j.1365-2966.2007.12487.x}, \href
  {https://ui.adsabs.harvard.edu/abs/2007MNRAS.382.1415S} {382, 1415}

\bibitem[\protect\citeauthoryear{{Schawinski} et~al.,}{{Schawinski}
  et~al.}{2009}]{Schawinski2009a}
{Schawinski} K.,  et~al., 2009, \mn@doi [\mnras]
  {10.1111/j.1365-2966.2009.14793.x}, \href
  {https://ui.adsabs.harvard.edu/abs/2009MNRAS.396..818S} {396, 818}

\bibitem[\protect\citeauthoryear{{Schawinski} et~al.,}{{Schawinski}
  et~al.}{2014}]{Schawinski2014}
{Schawinski} K.,  et~al., 2014, \mn@doi [\mnras] {10.1093/mnras/stu327}, \href
  {http://adsabs.harvard.edu/abs/2014MNRAS.440..889S} {440, 889}

\bibitem[\protect\citeauthoryear{{Schweizer}}{{Schweizer}}{1982}]{Schweizer1982}
{Schweizer} F.,  1982, \mn@doi [\apj] {10.1086/159573}, \href
  {https://ui.adsabs.harvard.edu/abs/1982ApJ...252..455S} {252, 455}

\bibitem[\protect\citeauthoryear{{Scoville} et~al.,}{{Scoville}
  et~al.}{2007}]{Scoville2007}
{Scoville} N.,  et~al., 2007, \mn@doi [\apjs] {10.1086/516585}, \href
  {https://ui.adsabs.harvard.edu/abs/2007ApJS..172....1S} {172, 1}

\bibitem[\protect\citeauthoryear{{Sengupta}, {Keel}, {Morrison}, {Windhorst},
  {Miller}  \& {Smith}}{{Sengupta} et~al.}{2022}]{Sengupta2022}
{Sengupta} A.,  {Keel} W.~C.,  {Morrison} G.,  {Windhorst} R.~A.,  {Miller} N.,
    {Smith} B.,  2022, \mn@doi [\apjs] {10.3847/1538-4365/ac3761}, \href
  {https://ui.adsabs.harvard.edu/abs/2022ApJS..258...32S} {258, 32}

\bibitem[\protect\citeauthoryear{{Song} et~al.,}{{Song}
  et~al.}{2021}]{Song2021}
{Song} H.,  et~al., 2021, \mn@doi [\mnras] {10.1093/mnras/staa3981}, \href
  {https://ui.adsabs.harvard.edu/abs/2021MNRAS.501.4635S} {501, 4635}

\bibitem[\protect\citeauthoryear{{Sousbie}}{{Sousbie}}{2011}]{Sousbie2011}
{Sousbie} T.,  2011, \mn@doi [\mnras] {10.1111/j.1365-2966.2011.18394.x}, \href
  {https://ui.adsabs.harvard.edu/abs/2011MNRAS.414..350S} {414, 350}

\bibitem[\protect\citeauthoryear{{Springel} \& {Hernquist}}{{Springel} \&
  {Hernquist}}{2005}]{Springel2005}
{Springel} V.,  {Hernquist} L.,  2005, \mn@doi [\apjl] {10.1086/429486}, \href
  {http://adsabs.harvard.edu/abs/2005ApJ...622L...9S} {622, L9}

\bibitem[\protect\citeauthoryear{{Stanford}, {Eisenhardt}  \&
  {Dickinson}}{{Stanford} et~al.}{1998}]{Stanford1998}
{Stanford} S.~A.,  {Eisenhardt} P.~R.,   {Dickinson} M.,  1998, \mn@doi [\apj]
  {10.1086/305050}, \href
  {https://ui.adsabs.harvard.edu/abs/1998ApJ...492..461S} {492, 461}

\bibitem[\protect\citeauthoryear{{Thomas}, {Maraston}, {Bender}  \& {Mendes de
  Oliveira}}{{Thomas} et~al.}{2005}]{Thomas2005}
{Thomas} D.,  {Maraston} C.,  {Bender} R.,   {Mendes de Oliveira} C.,  2005,
  \mn@doi [\apj] {10.1086/426932}, \href
  {https://ui.adsabs.harvard.edu/abs/2005ApJ...621..673T} {621, 673}

\bibitem[\protect\citeauthoryear{{Toomre} \& {Toomre}}{{Toomre} \&
  {Toomre}}{1972}]{Toomre1972}
{Toomre} A.,  {Toomre} J.,  1972, \mn@doi [\apj] {10.1086/151823}, \href
  {https://ui.adsabs.harvard.edu/abs/1972ApJ...178..623T} {178, 623}

\bibitem[\protect\citeauthoryear{{Watkins} et~al.,}{{Watkins}
  et~al.}{2022}]{Watkins2022}
{Watkins} A.~E.,  et~al., 2022, \mn@doi [\aap] {10.1051/0004-6361/202142627},
  \href {https://ui.adsabs.harvard.edu/abs/2022A&A...660A..69W} {660, A69}

\bibitem[\protect\citeauthoryear{{Weaver} et~al.,}{{Weaver}
  et~al.}{2022}]{Weaver2022}
{Weaver} J.~R.,  et~al., 2022, \mn@doi [\apjs] {10.3847/1538-4365/ac3078},
  \href {https://ui.adsabs.harvard.edu/abs/2022ApJS..258...11W} {258, 11}

\bibitem[\protect\citeauthoryear{{White} \& {Rees}}{{White} \&
  {Rees}}{1978}]{White1978}
{White} S.~D.~M.,  {Rees} M.~J.,  1978, \mn@doi [\mnras]
  {10.1093/mnras/183.3.341}, \href
  {https://ui.adsabs.harvard.edu/abs/1978MNRAS.183..341W} {183, 341}

\bibitem[\protect\citeauthoryear{{Yi} et~al.,}{{Yi} et~al.}{2005}]{Yi2005}
{Yi} S.~K.,  et~al., 2005, \mn@doi [\apjl] {10.1086/422811}, \href
  {https://ui.adsabs.harvard.edu/abs/2005ApJ...619L.111Y} {619, L111}

\bibitem[\protect\citeauthoryear{{Yoon}, {Park}, {Chung}  \& {Lane}}{{Yoon}
  et~al.}{2022}]{Yoon2022}
{Yoon} Y.,  {Park} C.,  {Chung} H.,   {Lane} R.~R.,  2022, \mn@doi [\apj]
  {10.3847/1538-4357/ac415d}, \href
  {https://ui.adsabs.harvard.edu/abs/2022ApJ...925..168Y} {925, 168}

\bibitem[\protect\citeauthoryear{{de Vaucouleurs}}{{de
  Vaucouleurs}}{1948}]{Vaucouleurs1948}
{de Vaucouleurs} G.,  1948, Journal des Observateurs, \href
  {https://ui.adsabs.harvard.edu/abs/1948JO.....31..113D} {31, 113}

\bibitem[\protect\citeauthoryear{{de Vaucouleurs}}{{de
  Vaucouleurs}}{1959}]{Vaucouleurs1959}
{de Vaucouleurs} G.,  1959, \mn@doi [Handbuch der Physik]
  {10.1007/978-3-642-45932-0_7}, \href
  {https://ui.adsabs.harvard.edu/abs/1959HDP....53..275D} {53, 275}

\bibitem[\protect\citeauthoryear{{de Vaucouleurs}}{{de
  Vaucouleurs}}{1977}]{Vaucouleurs1977}
{de Vaucouleurs} G.,  1977, in {Tinsley} B.~M.,  {Larson} Richard B.~Gehret
  D.~C.,  eds, Evolution of Galaxies and Stellar Populations. p.~43

\bibitem[\protect\citeauthoryear{{van Dokkum}, {Franx}, {Fabricant},
  {Illingworth}  \& {Kelson}}{{van Dokkum} et~al.}{2000}]{Dokkum2000}
{van Dokkum} P.~G.,  {Franx} M.,  {Fabricant} D.,  {Illingworth} G.~D.,
  {Kelson} D.~D.,  2000, \mn@doi [\apj] {10.1086/309402}, \href
  {https://ui.adsabs.harvard.edu/abs/2000ApJ...541...95V} {541, 95}

\makeatother
\end{thebibliography}



\appendix

\section{Density maps}
\label{app:dens_maps}

In this section we present all density maps used in our study (see text in Section \ref{sec:disperse} for details about their construction). Note that the density map for the redshift range $0.218<z<0.243$ is shown in Figure \ref{fig:EXMPslice}. 

\begin{figure*}
    \centering
    \includegraphics[width=0.45\textwidth]{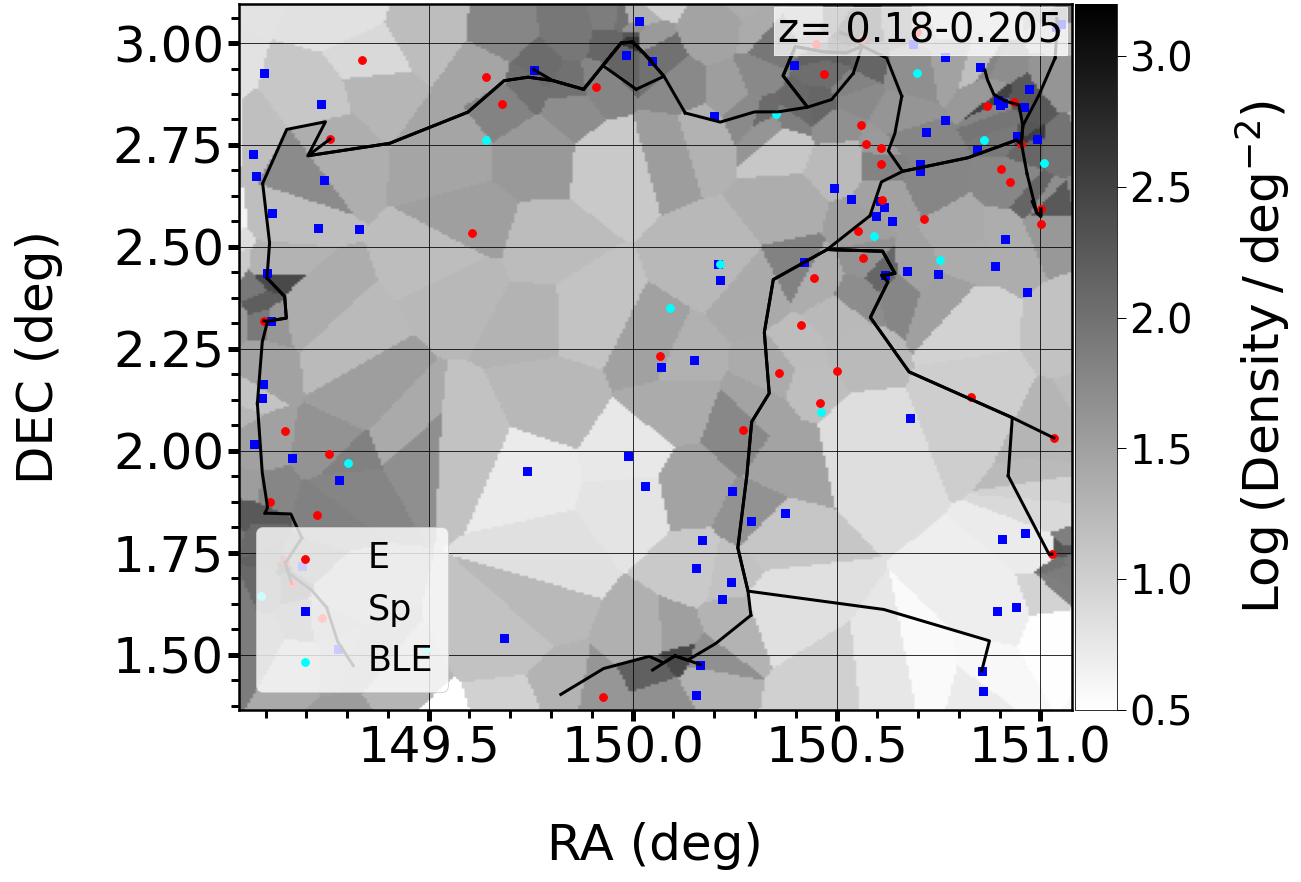} 
    \includegraphics[width=0.45\textwidth]{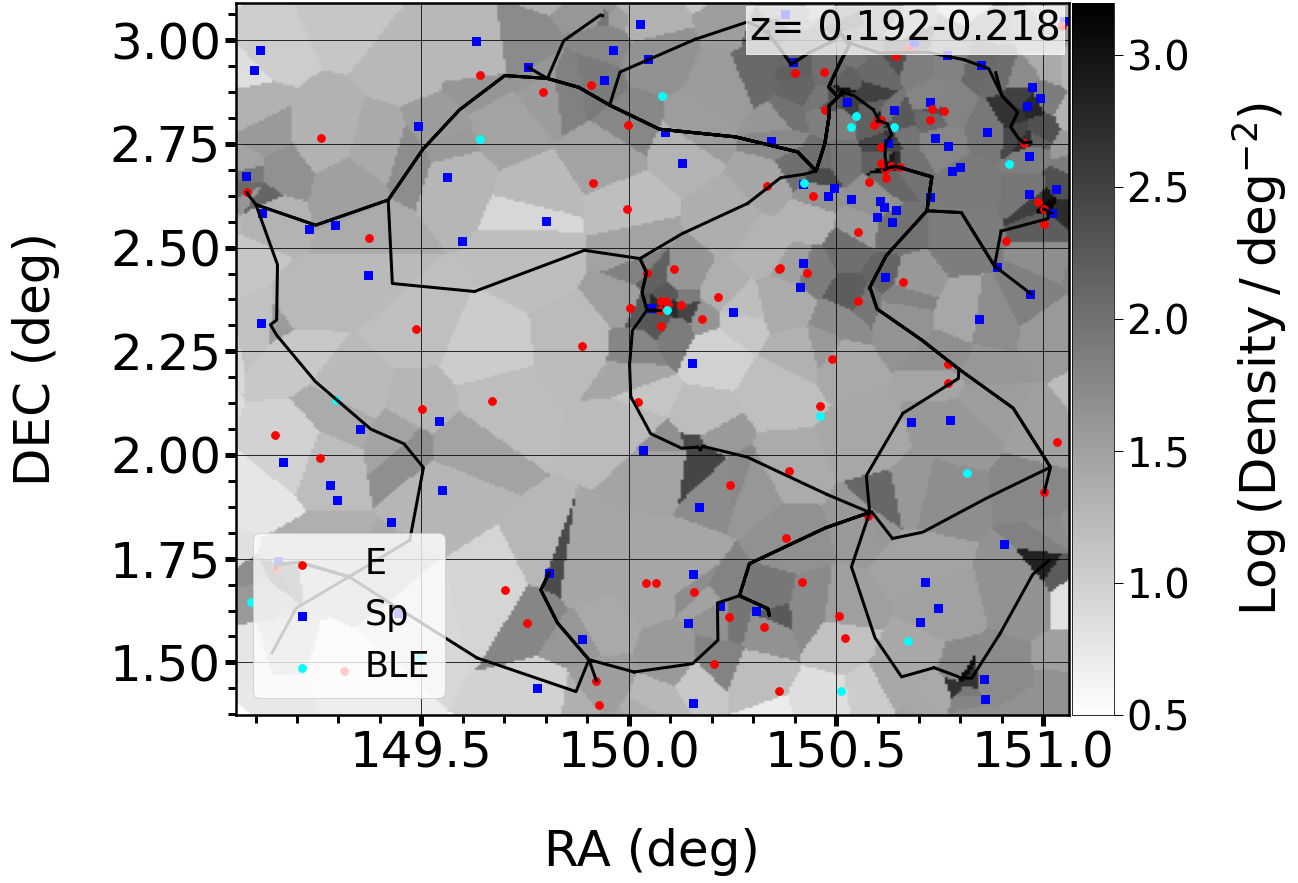}
    \includegraphics[width=0.45\textwidth]{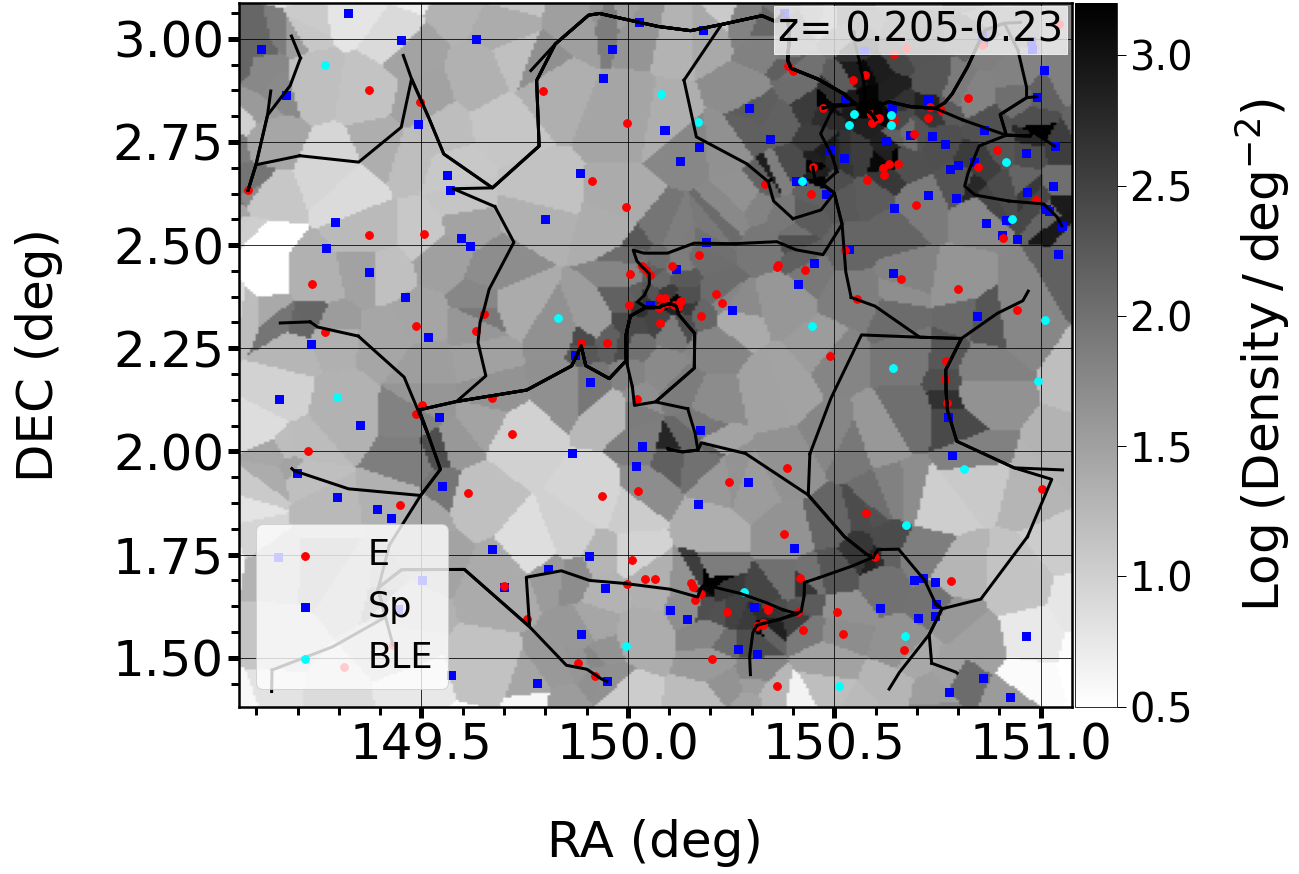}
    \includegraphics[width=0.45\textwidth]{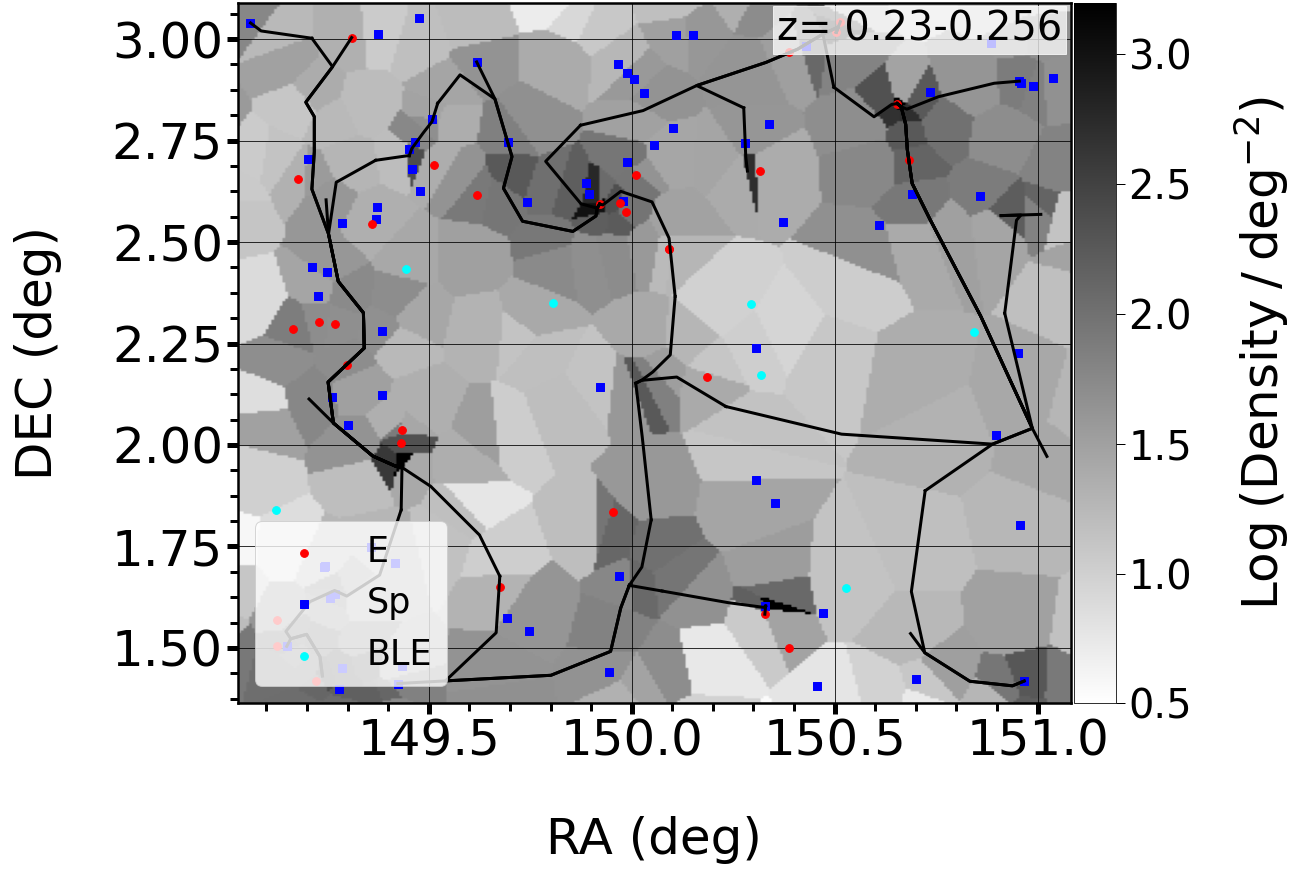}
    \includegraphics[width=0.45\textwidth]{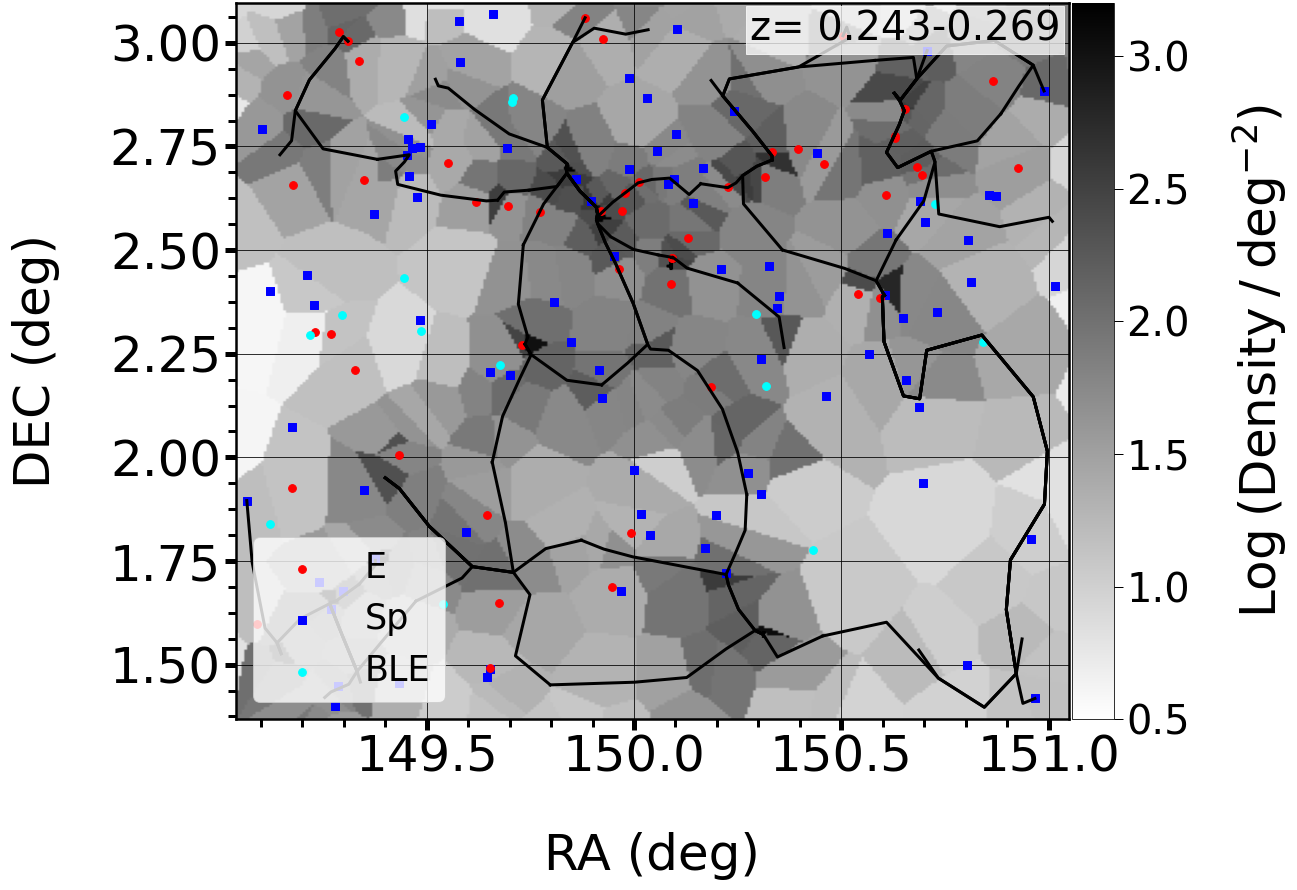}
    \includegraphics[width=0.45\textwidth]{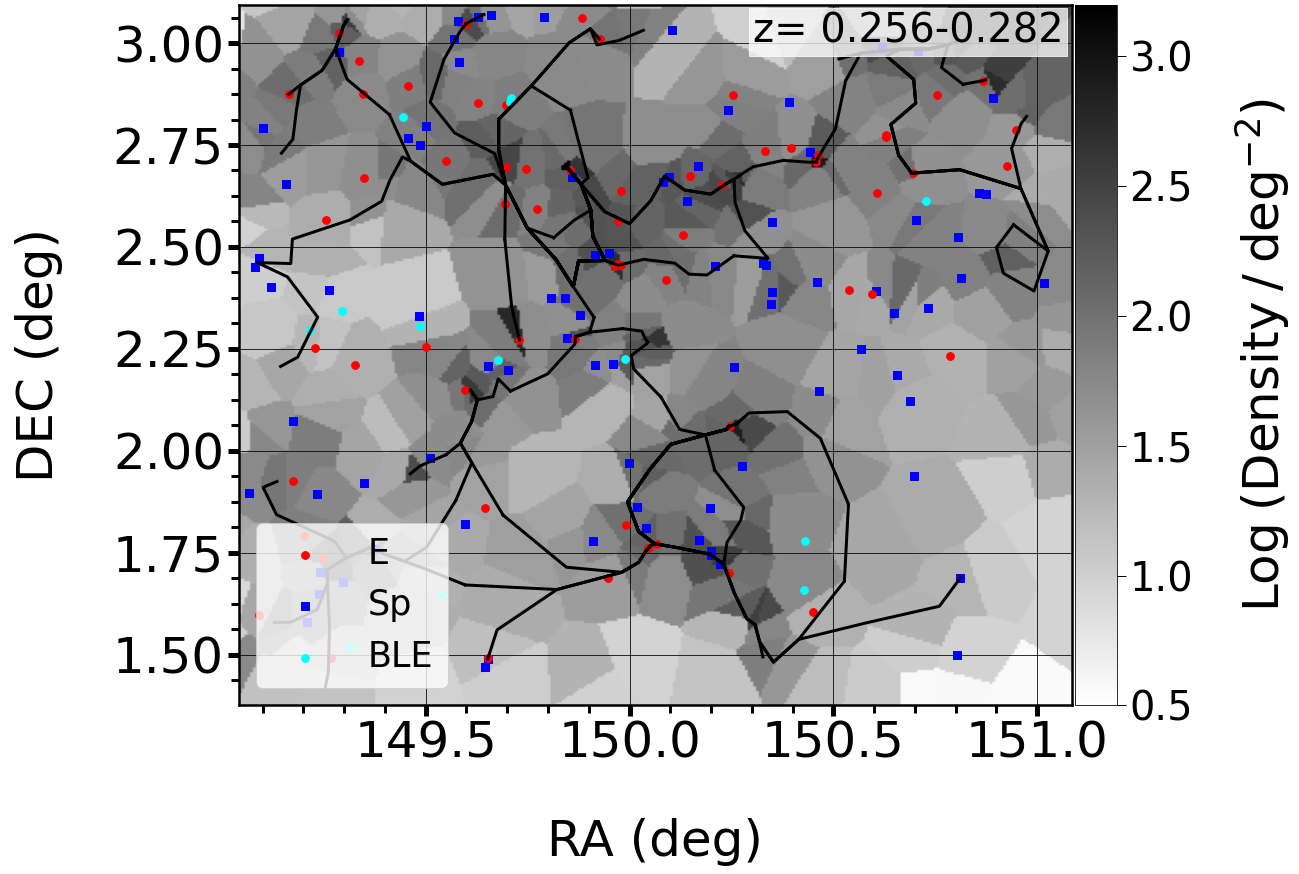}
    \includegraphics[width=0.45\textwidth]{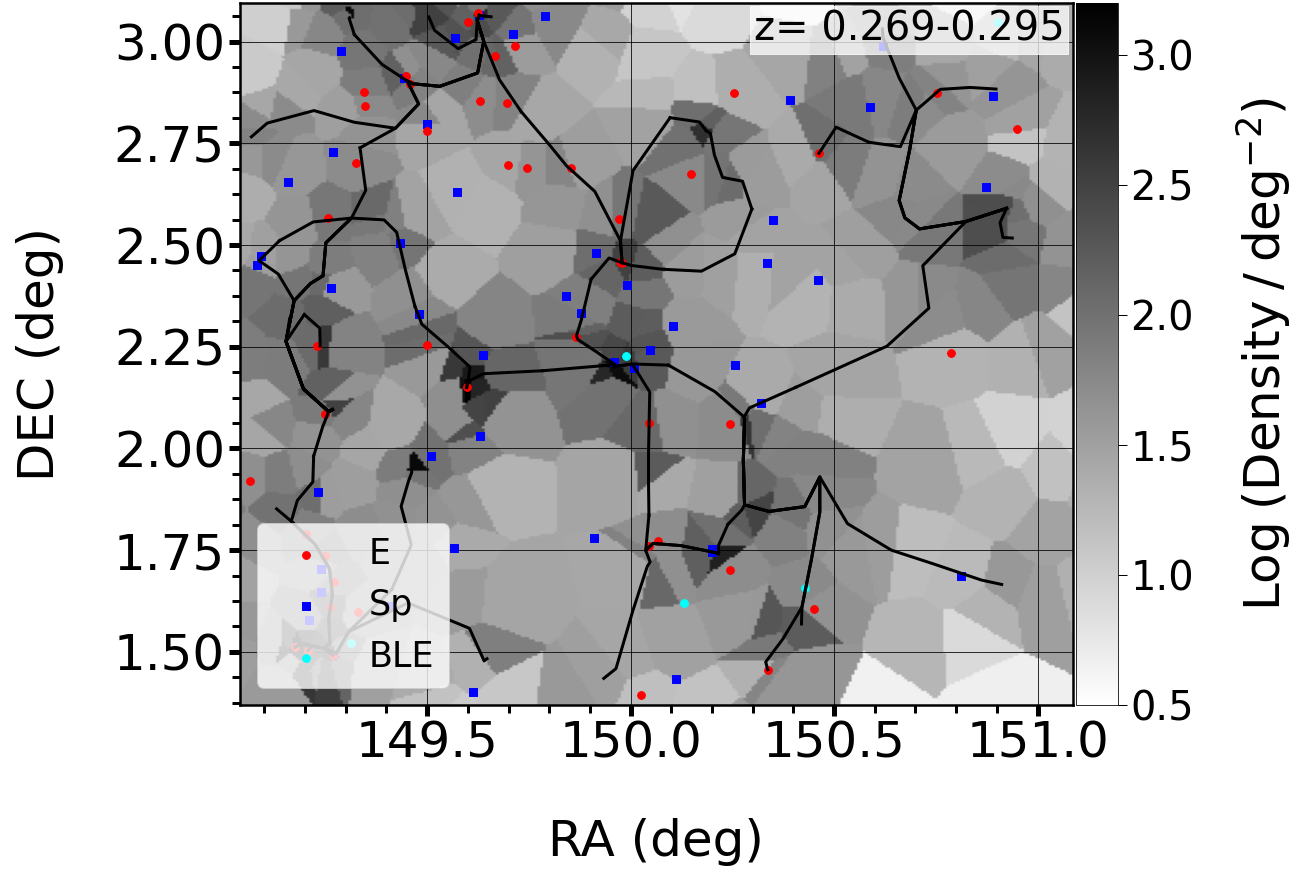}
    \includegraphics[width=0.45\textwidth]{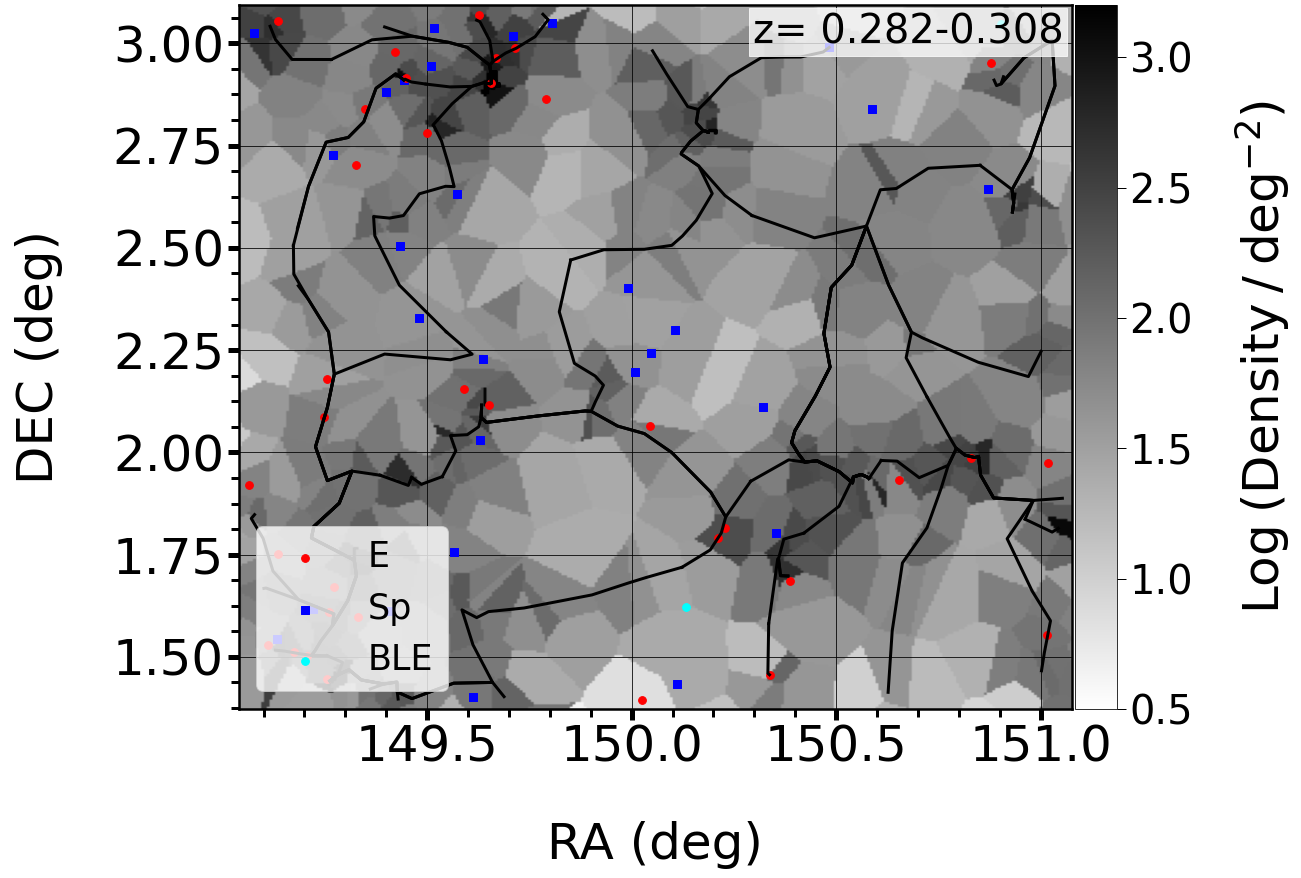}
    \caption{Density maps and skeletons (shown using black solid lines) used in this study. Different morphological classes (red: ellipticals, blue: spirals, cyan: blue ellipticals) are shown overlaid.}
    \label{fig:slice_1_z=0.18-0.205}
\end{figure*}

\clearpage


\section{Examples of sersic fits}
\label{app:sersic}

\begin{figure}
    \centering
    \includegraphics[width=0.45\textwidth]{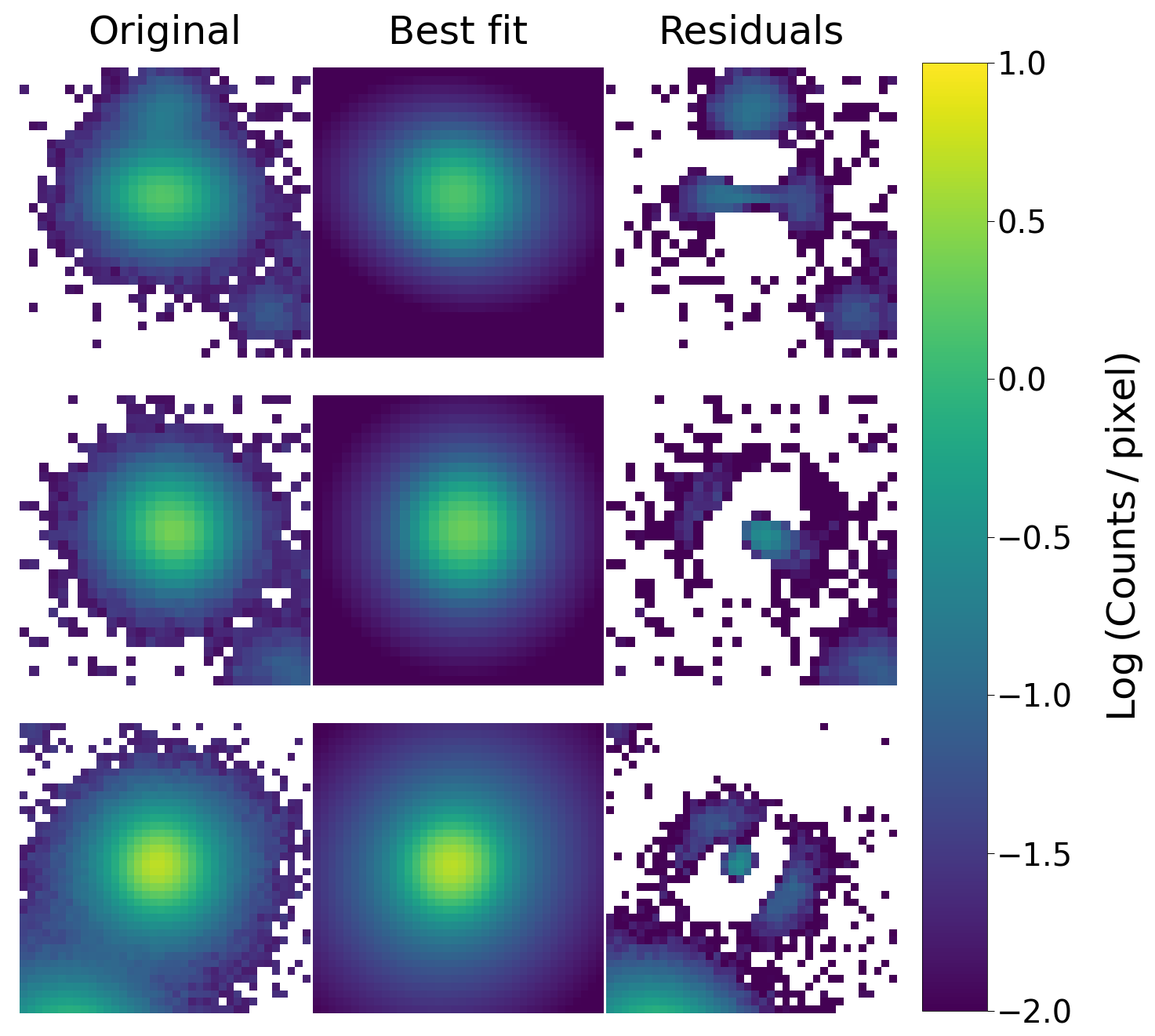}
    \caption{Sersic fits for three random blue ellipticals which have spectroscopic redshifts available. From top to bottom the fitted Sersic indices and maximum residual values are 4.56, 7, 5.13 and 0.1, 0.3 and 0.7 counts pixel$^{-1}$ respectively.}
    \label{fig:specz}
\end{figure}

\begin{figure}
    \centering
    \includegraphics[width=0.45\textwidth]{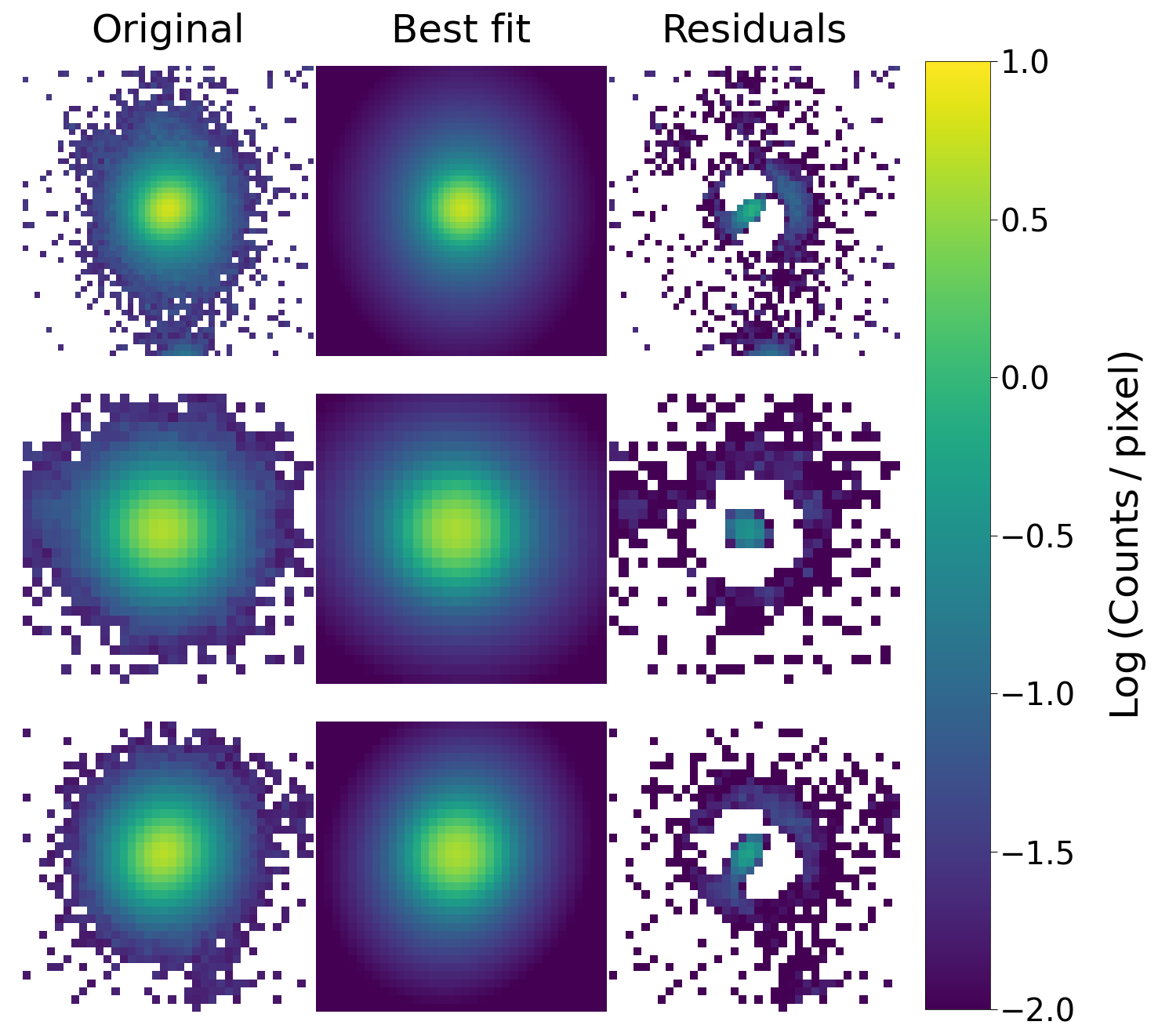}
    \caption{Sersic fits for three blue ellipticals which have only photometric redshifts available. From top to bottom the fitted Sersic indices and maximum residual values are 7, 6.5, 4.4 and 0.8, 0.3 and 0.5 counts pixel$^{-1}$ respectively.}
    \label{fig:photz}
\end{figure}

\begin{figure}
    \centering
    \includegraphics[width=0.45\textwidth]{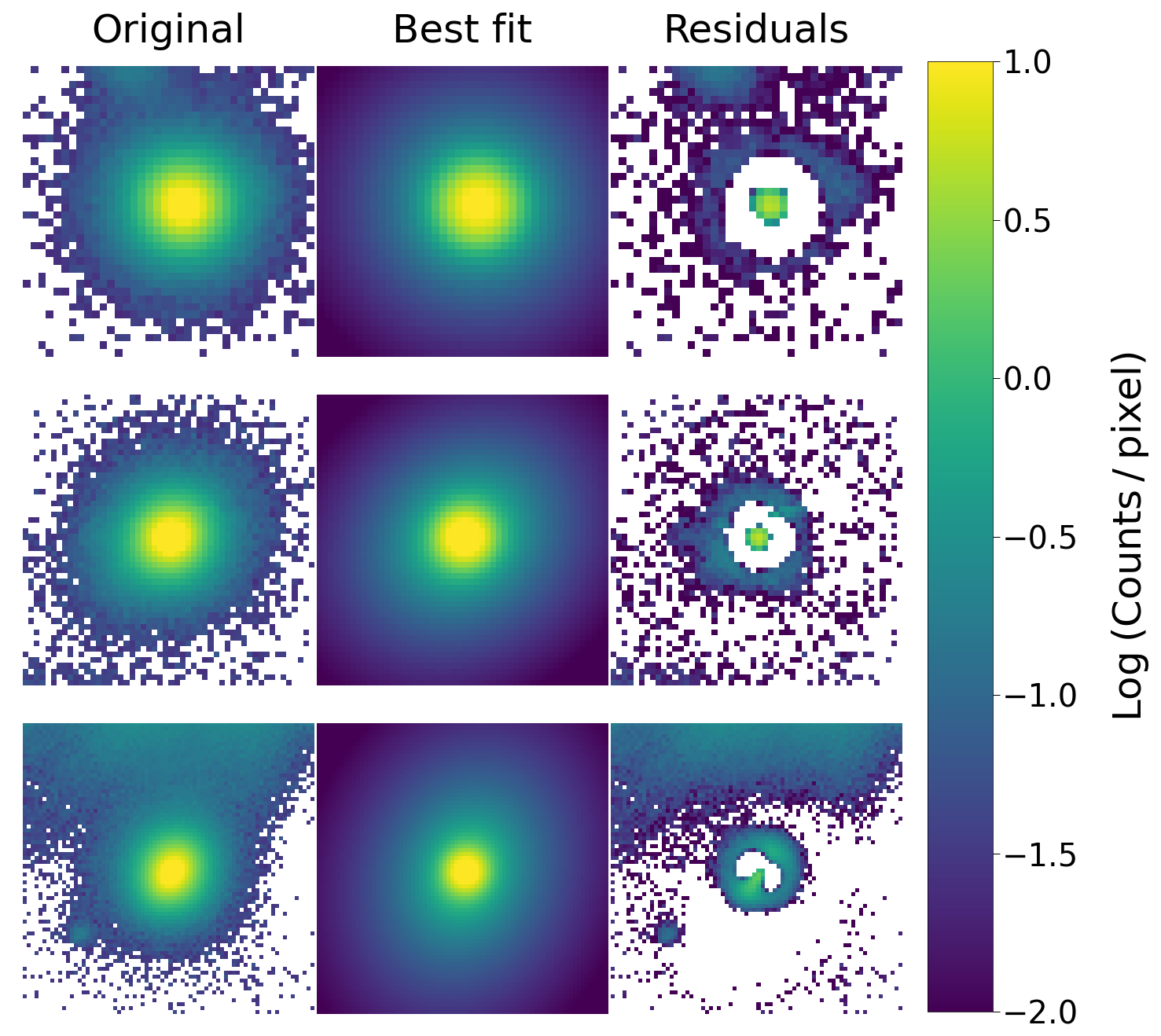}
    \caption{Sersic fits for three objects classified as stars by our method. From top to bottom the fitted Sersic indices and maximum residual values are 7, 5, 6.1 and 4.7, 6.1 and 1.9 counts pixel$^{-1}$ respectively.}
    \label{fig:star}
\end{figure}

In this section we show Sersic fits for typical examples of blue ellipticals which have both photometric and spectroscopic redshifts (Figure \ref{fig:specz}), blue ellipticals which have only photometric redshifts available (Figure \ref{fig:photz}) and objects classified as stars (Figure \ref{fig:star}) by our method. 


\bsp	
\label{lastpage}
\end{document}